\documentclass[%
 reprint,%
 amsmath,%
 amssymb,%
 aps,%
 floatfix,
 superscriptaddress]{revtex4-2}

\usepackage{graphicx}
\graphicspath{{./figs}}
\usepackage{mathtools}
\usepackage{amsmath, amssymb}
\usepackage{dsfont}
\usepackage{bm}
\usepackage[dvipsnames]{xcolor}
\usepackage[colorlinks,%
            pdfusetitle,%
            bookmarksopen=false,%
            urlcolor=Blue,%
            citecolor=Blue,%
            linkcolor=Blue,%
            pdfencoding=auto,%
            psdextra]{hyperref}
\usepackage{physics}
\usepackage{upgreek}
\usepackage{bbm}
\usepackage[normalem]{ulem}
\usepackage{siunitx}
\usepackage{booktabs}

\newcommand{\Sr}{\ensuremath{{}^\text{88}\text{Sr}} }

\newcommand{\trippzero}{\ensuremath{{}^\text{3} \mathrm{P}_\text{0}} }
\newcommand{\trippone}{\ensuremath{{}^\text{3} \mathrm{P}_\text{1}} }
\newcommand{\tripptwo}{\ensuremath{{}^\text{3} \mathrm{P}_\text{2}} }
\newcommand{\singszero}{\ensuremath{{}^\text{1} \mathrm{S}_\text{0}} }
\newcommand{\singpone}{\ensuremath{{}^\text{1} \mathrm{P}_\text{1}} }
\newcommand{\tripsone}{\ensuremath{{}^\text{3} \mathrm{S}_\text{1}} }

\newcommand{\Tgate}{\ensuremath{T_G}}
\newcommand{\UGHZ}{\ensuremath{\hat{\mathcal{U}}}}

\newcommand{\Nmax}{\ensuremath{N_{\rm max}}}
\newcommand{\parityz}{\ensuremath{\hat{\mathcal{P}}_z}}

\newcommand{\wanniernR}{\ensuremath{\ket{w_{n, \vec{R}}}}}

\newcommand{\blochnq}{\ensuremath{\ket{\psi_{n,\vec{q}}}}}

\newcommand{\Fclock}{0.9962(7)}

\newcommand{\Frydspam}{0.995(2)}

\newcommand{\Frydbspam}{0.986(3)}

\newcommand{\Fbellraw}{0.983(2)}
\newcommand{\bellpop}{0.990(2)}
\newcommand{\bellpar}{0.975(3)}

\newcommand{\adevfour}{\ensuremath{1.81(3) \times 10^{-14}/\sqrt{\tau}}}
\newcommand{\gainfourrelcss}{\SI{3.9(2)}{dB}}
\newcommand{\gainfourrelsql}{\SI{2.1(1)}{dB}}
\newcommand{\cssfourrelsql}{\SI{1.8(1)}{dB}}

\renewcommand{\figurename}{Fig.}

\begin{document}


\title{Multi-qubit gates and Schrödinger cat states in an optical clock}

\author{Alec Cao}
\author{William J. Eckner}
\author{Theodor \surname{Lukin Yelin}}
\author{Aaron W. Young}

\affiliation{%
JILA, University of Colorado and National Institute of Standards and Technology,
and Department of Physics, University of Colorado, Boulder, Colorado 80309, USA
}%

\author{Sven Jandura}
\affiliation{University of Strasbourg and CNRS, CESQ and ISIS (UMR 7006), aQCess, 67000 Strasbourg, France}

\author{Lingfeng Yan}
\author{Kyungtae Kim}
\affiliation{%
JILA, University of Colorado and National Institute of Standards and Technology,
and Department of Physics, University of Colorado, Boulder, Colorado 80309, USA
}%

\author{Guido Pupillo}
\affiliation{University of Strasbourg and CNRS, CESQ and ISIS (UMR 7006), aQCess, 67000 Strasbourg, France}

\author{Jun Ye}
\author{Nelson \surname{Darkwah Oppong}}
\author{Adam M. Kaufman}
\email[e-mail:$\,$]{adam.kaufman@colorado.edu}
\affiliation{%
JILA, University of Colorado and National Institute of Standards and Technology,
and Department of Physics, University of Colorado, Boulder, Colorado 80309, USA
}%

\date{\today}

\begin{abstract}
    Many-particle entanglement is a key resource for achieving the fundamental precision limits of a quantum sensor~\cite{pezze2018quantum}. Optical atomic clocks~\cite{ludlow2015optical}, the current state-of-the-art in frequency precision, are a rapidly emerging area of focus for entanglement-enhanced metrology~\cite{colombo2022entanglement,pedrozo2020entanglement, robinson2022direct, eckner2023realizing}. Augmenting tweezer-based clocks featuring microscopic control and detection~\cite{norcia2019seconds,madjarov2019atomic, young2020half,shaw2024multi} with the high-fidelity entangling gates developed for atom-array information processing~\cite{evered2023high, ma2023high} offers a promising route towards leveraging highly entangled quantum states for improved optical clocks. Here we develop and employ a family of multi-qubit Rydberg gates to generate Schrödinger cat states of the Greenberger-Horne-Zeilinger (GHZ) type with up to 9 optical clock qubits in a programmable atom array. In an atom-laser comparison at sufficiently short dark times, we demonstrate a fractional frequency instability below the standard quantum limit using GHZ states of up to 4 qubits. However, due to their reduced dynamic range, GHZ states of a single size fail to improve the achievable clock precision at the optimal dark time compared to unentangled atoms~\cite{huelga1997improvement}. Towards overcoming this hurdle, we simultaneously prepare a cascade of varying-size GHZ states to perform unambiguous phase estimation over an extended interval~\cite{higgins2009demonstrating,berry2009perform,kessler2014heisenberg,komar2014quantum}. These results demonstrate key building blocks for approaching Heisenberg-limited scaling of optical atomic clock precision. 
\end{abstract}

\maketitle

Quantum systems have revolutionized sensing and measurement technologies~\cite{degen2017quantum}, spanning applications from nanoscale imaging with nitrogen vacancy centers~\cite{schirhagl2014nitrogen} to gravimetry with atom interferometers~\cite{bongs2019taking}, and timekeeping based on optical atomic clocks~\cite{ludlow2015optical}. A major precision barrier for such devices is the quantum projection noise (QPN) arising from inherently probabilistic quantum measurements. Because of QPN, a measurement on $N$ independent and identical quantum sensors will have an uncertainty scaling as $1/\sqrt{N}$, known as the standard quantum limit (SQL). However, the fundamental precision bound given by quantum theory is the Heisenberg limit (HL) with $1/N$ scaling for linear observables. Improving measurements from the SQL towards the HL using entangled or non-classical resources is the central thrust of quantum-enhanced metrology~\cite{pezze2018quantum}, an approach which has already yielded benefits in fundamental physics~\cite{tse2019quantum, backes2021quantum} and biology~\cite{casacio2021quantum}.

The intersection of programmable atom arrays with optical atomic clocks provides a novel opportunity in this endeavor. The former have emerged as one of the leading architectures for quantum information processing~\cite{bluvstein2022quantum,graham2022multi,bluvstein2024logical}, with advances in Rydberg-gate design~\cite{jandura2022time,levine2019parallel} now enabling controlled-phase (CZ) gate fidelites as high as 0.995~\cite{evered2023high,ma2023high}. The latter now routinely achieve fractional frequency uncertainties at or below the $10^{-18}$ level~\cite{bloom2014optical,ushijima2015cryogenic,mcgrew2018atomic,Brewer2019Quantum, oelker2019demonstration,bothwell2022resolving, zheng2022differential}, with synchronous comparisons allowing for stability near or at the SQL. Merging these capabilities becomes possible with tweezer-controlled optical atomic clocks~\cite{norcia2019seconds, madjarov2019atomic}, which have demonstrated a relative instability of $5 \times 10^{-17}/\sqrt{\tau}$~\cite{young2020half} (where $\tau$ denotes the averaging time in seconds). The integration of high-fidelity entangling gates for generating metrologically useful many-body states in a clock-qubit atom array~\cite{schine2022long,scholl2023erasure2} serves as a natural path towards entanglement-enhanced measurements at the precision frontier.

Of particular interest is the generation and use of Schrödinger cats, coherent superpositions of two macroscopically distinct quantum states~\cite{frowis2012measures}. Specifically, the maximally entangled GHZ-type cat state of $N$ qubits
\begin{align}
    \ket{\mathrm{GHZ}} = \frac{1}{\sqrt{2}} \left( \ket{0}^{\otimes N} + \ket{1}^{\otimes N} \right),
    \label{eqghz}
\end{align}
accumulates phase $N$-times faster than unentangled qubits and saturates the HL~\cite{toth2014quantum}. However, GHZ states also suffer from increased sensitivity to dephasing noise~\cite{huelga1997improvement} and fragility to decay and loss, making them difficult to create and use. This delicate nature is a core reason that large GHZ-state production has become a standard benchmark for quantum processors~\cite{pogorelov2021compact, moses2023race, bao2024schrodinger}. On the other hand, quantum metrology faces the key question of whether such fragility compromises the practical utility of these states. A growing number of small-scale demonstrations suggest that GHZ states can indeed perform below the SQL in a broad range of contexts~\cite{leibfried2004toward,nagata2007beating,jones2009magnetic,facon2016sensitive}, though their application to clock operation has remained largely unexplored in experiments.

\begin{figure*}
    \centering
    \includegraphics[width=\textwidth]{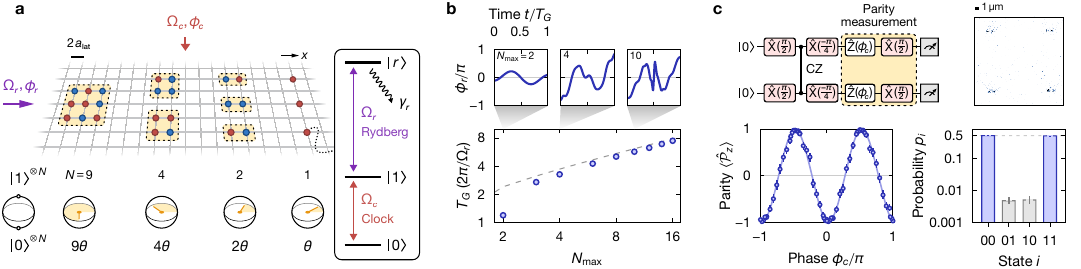}
    \caption{\label{fig1}%
    \textbf{Global multi-qubit gates in a ${}^\text{88}\text{Sr}$ atom array.}
    \textbf{a}, Schematic of the experimental setup.
    ${}^\text{88}\text{Sr}$ atoms in the states~$\ket{0}$ (blue circles) and~$\ket{1}$ (red circles) are arranged into different ensemble sizes $N$ in an optical lattice (gray lines, lattice spacing $a_\mathrm{lat} \approx \SI{575}{\nano\meter}$).
    The optical $\ket{0}\leftrightarrow\ket{1}$ transition is driven by a clock laser (red arrow).
    To generate entanglement, a separate laser (purple arrow) globally couples~$\ket{1}$ to a high-lying Rydberg state~$\ket{r}$ with decay rate~$\gamma_r$ (see level diagram).
    This Rydberg laser realizes a global multi-qubit gate~$\UGHZ$ [see Eq.~\eqref{eq:UGHZ}] that simultaneously produces states of the form $\ket{\mathrm{GHZ}}$ [up to a global $Z$ rotation, see Eq.~\eqref{eqghz} and Methods] with variable size~$N$ (yellow shaded areas).
    In a Ramsey sequence, $\ket{\mathrm{GHZ}}$ acquires $N$-times the single-particle phase $\theta$ (bottom Bloch spheres).
    \textbf{b}, Numerically determined duration $T_G$ (blue circles) for the time-optimal multi-qubit gate~$\UGHZ$ with variable maximum GHZ-state size~$\Nmax$ (bottom panel).
    The dashed line shows the fitted scaling $\Tgate \sim \Nmax^{0.59}$.
    The top panels show the time-dependent Rydberg laser phase~$\phi_r$ for~$\Nmax=2,4,10$.
    \textbf{c}, Generation of Bell states using the time-optimal CZ gate implementation from Ref.~\cite{evered2023high}.
    The top row shows (left) the equivalent quantum circuit used for preparing and characterizing Bell states and (right) a single-shot fluorescence image of the atomic pairs.
    The bottom row shows (left) the expectation of the parity $\parityz$ (blue circles) for variable phase~$\phi_c$ and (right) the probability~$p_i$ (bar graph) to observe the two-particle state~$i$.
    A sinusoidal fit to the parity (light-blue line) yields the contrast $C=\bellpar$.
    With $p_{00} + p_{11} = \bellpop$, this corresponds to a raw Bell-state fidelity of~$\mathcal{F}_{\rm raw}=\Fbellraw$.}
\end{figure*}

In this article, we experimentally investigate the generation and metrological performance of GHZ states in an array of strontium clock qubits. These explorations mark the first realization of GHZ states in a neutral-atom optical clock, as well as the first time that GHZ states have been used for below-SQL performance in an atom-laser comparison (with a restricted dark time). Underlying these results is the extension of the time-optimal Rydberg-gate toolkit~\cite{jandura2022time,evered2023high} to a class of multi-qubit operations for producing fully connected graph states. Using these gates, we realize a raw Bell-state fidelity of $\Fbellraw$ and create GHZ states of up to 9 atoms. In an atom-laser frequency comparison, an instability below the SQL (at the $10^{-14}/\sqrt{\tau}$ level) is achieved at a fixed and sufficiently short Ramsey dark time of $\SI{3}{\milli \second}$ for GHZ states of up to 4 atoms. Towards overcoming the dark time restriction, we explore multi-ensemble metrology with simultaneously prepared GHZ states of varying size to recover unambiguous phase estimation over a range comparable to unentangled atoms.

\subsection*{Gate design and GHZ-state preparation}

Our experiment features a \Sr atom array trapped in an optical lattice and programmably rearranged by optical tweezers~\cite{eckner2023realizing}. The qubits are encoded on the optical transition comprised of the ground \singszero state ($\ket{0}$) and clock \trippzero state ($\ket{1}$). Global single-qubit $\hat{X}(\theta)$ rotations are implemented by clock laser pulses with a typical Rabi frequency of $\Omega_c = 2 \pi \times \SI{300}{\hertz}$, and global $\hat{Z}(\theta)$ rotations by changing the clock laser phase; here $\theta$ denotes the angle of rotation (see Methods). Entanglement is generated by globally coupling $\ket{1}$ to the 47s \tripsone Rydberg state ($\ket{r}$), with typical Rabi frequencies in the range $\Omega_r = 2\pi \times $3--4\,MHz. High-fidelity clock and Rydberg operations are a key enabling feature of this work (see Extended Data Fig.~\ref{fig2ed} and Methods).

Each experiment begins with atoms arranged into small, isolated ensembles (see Fig.~\ref{fig1}a). Starting from $\ket{0}^{\otimes N}$, an $\hat{X}(\pi/2)$ rotation initializes all atoms into an equal superposition of $\ket{0}$ and $\ket{1}$. We then turn on the Rydberg coupling, during which strong Rydberg interactions (compared to $\Omega_r$) suppress multiple excitations to $\ket{r}$ within an ensemble. This Rydberg blockade effect causes collective oscillations with a $\sqrt{n}$-enhanced Rabi frequency for states with $n$ atoms in $\ket{1}$~\cite{lukin2001dipole,urban2009observation,dudin2012observation}; explicitly, $\hat{n}=\sum_{j=1}^N \hat{n}_j$ with $\hat{n}_j = \ket{1_j} \bra{1_j}$ for an $N$-atom ensemble indexed by $j$. By modulating the Rydberg laser phase $\phi_r$ in time, the blockaded Rabi oscillations for different $n$ can be steered to simultaneously return to the computational subspace while acquiring an $n$-dependent phase. This is the core mechanism underlying many recent implementations of Rydberg logic gates~\cite{levine2019parallel, jandura2022time,evered2023high, ma2023high}. 

Here we apply optimal control to $\phi_r$ (see Methods) to implement the multi-qubit gate
\begin{align}
    \UGHZ = \exp \left( i \frac{\pi}{2} \hat{n}^2 \right).
    \label{eq:UGHZ}
\end{align}
Up to a global phase and $\hat{Z}(-\pi/2)$ rotation, $\UGHZ$ applies a CZ gate to every pair of qubits~(see Methods). When $\UGHZ$ is applied to $\hat{X}(\pi/2) \ket{0}^{\otimes N}$, the fully connected graph state is produced, which connects to a GHZ state under a global $\hat{X}(\pi/2)$ rotation (see Methods). Illustrative examples of $\phi_r$ are shown in Fig.~\ref{fig1}b, with each implementing $\UGHZ$ for any ensemble size $N \leq \Nmax$. The gate duration $\Tgate$ increases only sublinearly in $\Nmax$, potentially improving fidelities compared to a standard GHZ-state preparation circuit, with $N-1$ two-qubit gates, when Rydberg decay is the dominant error. We note that due to the finite range of the Rydberg blockade, multi-qubit Rydberg gates are most practical for an intermediate range of $N$.

The GHZ-state fidelity of a state is the maximal overlap with $\ket{\mathrm{GHZ}}$ [Eq.~\eqref{eqghz}] under a global $\hat{Z}$ rotation (see Methods). It can be obtained by measuring the populations in $\ket{0}^{\otimes N}$ and $\ket{1}^{\otimes N}$, in addition to the contrast of a parity oscillation which characterizes the coherence (see Methods); the $N$-qubit parity $\parityz = (-1)^N e^{i \pi \hat{n}}$ has eigenvalues $\mathcal{P}_z = +1$ ($-1$) for even (odd) $N-n$. Before implementing $\UGHZ$, we benchmark our system by measuring this fidelity for a two-qubit Bell state $\left( \ket{00} + \ket{11} \right)/\sqrt{2}$, which corresponds to $\ket{\mathrm{GHZ}}$ in Eq.~\eqref{eqghz} for $N=2$. The Bell state is generated by applying an $\hat{X}(-\pi/4)$ rotation after a CZ gate (see Fig.~\ref{fig1}c), and we use the CZ gate implementation with sinusoidal $\phi_r$ described in Ref.~\cite{evered2023high}. We achieve a raw Bell-state fidelity of $\mathcal{F}_{\rm raw}=\Fbellraw$ (see Fig.~\ref{fig1}c). This significantly improves $\mathcal{F}_{\rm raw}=0.871(16)$ reported in our previous work using adiabatic dressing gates~\cite{schine2022long}, and is comparable to the best achieved in neutral atoms on the alkali hyperfine qubit~\cite{evered2023high}.

\begin{figure}
    \centering
    \includegraphics[width=\columnwidth]{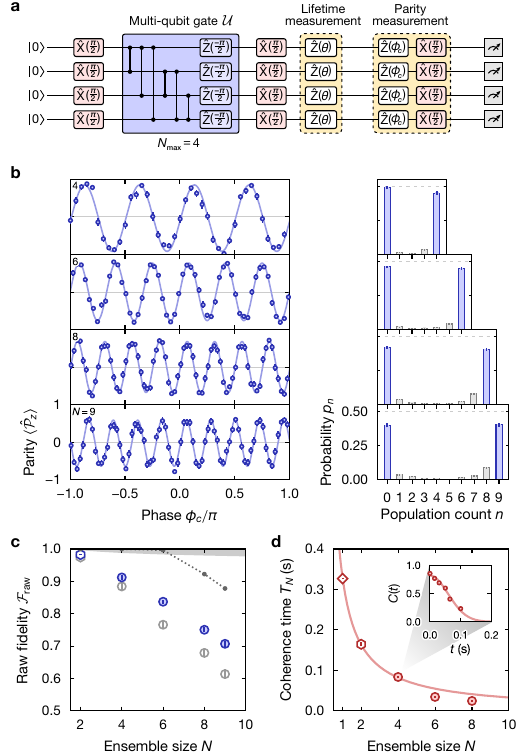}
    \caption{\label{fig2}%
    \textbf{Preparing $N$-particle GHZ states with multi-qubit gates.}
    \textbf{a}, Equivalent quantum circuit for preparing and characterizing $N$-particle GHZ states (shown for~$N=N_\mathrm{max} = 4$).
    The blue box represents the multi-qubit gate~$\UGHZ$ in terms of CZ gates (see Methods).
    \textbf{b}, Measurement of the $N$-particle parity (blue circles, left column) with sinusoidal fits (light-blue line) and the probability~$p_n$ to observe an $N$-particle state with the population count~$n$ (bar graph, right column).
    \textbf{c}, Raw GHZ-state fidelity (filled blue circles) for variable ensemble size~$N$ using $\mathcal{F}_{\rm raw} = (C + p_0 + p_N)/2$ with parity contrast $C$ determined from the fits in panel~b (shown as gray markers).
    The empty hexagon marker corresponds to the two-particle Bell state (Fig.~\ref{fig1}c).
    The gray shaded area shows an approximate upper bound due to finite Rydberg state lifetime (see Methods).
    The gray points and dotted line show the simulated fidelity only taking into account the finite Rydberg blockade (see Extended Data Table~\ref{tabed1} and Methods).
    \textbf{d},
    Coherence time of the $N$-particle GHZ states (filled circles) extracted from the parity contrast~$C$ after variable hold time~$t$ (see panel~a).
    During~$t$, the fluctuating atom-laser detuning $\delta(t)$ is integrated into a random phase $\theta$ (see Methods), causing a rotation $\hat{Z}(\theta)$ (panel~a).
    The light-red line corresponds to the scaling~$T_1/N$.
    The empty hexagon marker shows the lifetime for the two-particle Bell state and the empty diamond marker shows the single-particle lifetime without applying~$\mathcal{U}$.
    The inset shows the $N=4$ data from which we extract the lifetime using a Gaussian fit (light-red line).
    }
\end{figure}

Next we apply the GRAPE-optimized form of $\phi_r$ to implement $\UGHZ$ and produce $N>2$ GHZ states. The GHZ state is generated by an $\hat{X}(\pi/2)$ rotation after applying $\UGHZ$ (see Fig.~\ref{fig2}a and Methods). For each $N$, we use $\phi_r$ for $\Nmax = N$ (except for $N=9$, in which we use $\Nmax = 10$). The fidelity is again extracted through populations and parity contrast measurements (see Fig.~\ref{fig2}b and Methods). A summary of the raw fidelities is plotted in Fig.~\ref{fig2}c. We also show the raw parity contrasts, which bound the GHZ-state fidelity from below and are the figure of merit in metrology applications. The contrasts are $>0.6$ for all $N\leq9$, certifying genuine 9-particle entanglement~\cite{sackett2000experimental}. The fidelities corrected for measurement errors are all comparable to the raw values (see Extended Data Table~\ref{tabed2} and Methods). While larger neutral-atom GHZ states have been produced on a short-lived Rydberg qubit~\cite{omran2019generation}, these results represent the largest GHZ states to be created on a long-lived neutral-atom qubit, with fidelities on par with or better than the previous state-of-the-art~\cite{graham2022multi,evered2023high}.

The observed raw fidelities decrease approximately linearly in $N$, in contrast to the sublinear expectation based solely on Rydberg decay (gray shaded area, see Fig.~\ref{fig1}b and Methods). A major challenge for multi-qubit Rydberg gates is the finite, spatially decaying interaction. The expected fidelities accounting for this effect (gray points with dotted lines) are obtained from simulations including the varying pair energies assuming a $1/r^6$ scaling, with $r$ the atomic separation; we note that this scaling can break down for the closest atomic spacings used (see Methods and Extended Data Table~\ref{tabed1}). The expected blockade violation (multiple Rydberg excitations) is $\lesssim 10^{-4}$ for the $N$ shown, suggesting that the infidelity mainly arises from inhomogeneous effective Stark shifts~\cite{jandura2022time}. The finite blockade limits the maximum GHZ-state size achieved along with technical restrictions on our atom rearrangement (see Methods); mitigating this effect by reducing $\Omega_r$ is challenging due to recapture loss (see Extended Data Fig.~\ref{fig3ed} and Methods). Various other errors sources are considered in the Methods and Extended Data Fig.~\ref{fig4ed}.

To characterize the coherence time of the GHZ states, we repeat the parity contrast measurements with a variable hold time $t$ before the parity analysis $\hat{X}(\pi/2)$ rotation (see Fig.~\ref{fig2}d). A Gaussian decay of coherence is observed, indicating inhomogeneous broadening that we attribute to magnetic field noise (see Methods). Under correlated, non-Markovian noise, the GHZ-state coherence time is expected to obey $T_N = T_1/N$~\cite{monz201114,graham2022multi}, where $T_1$ is the coherence time for unentangled atoms. This behavior is observed for $N \leq 4$, but the data for $N=6$ and 8 show a reduction relative to this. 

\subsection*{GHZ-state atom-laser comparison}

\begin{figure}[!]
    \centering
    \includegraphics[width=\columnwidth]{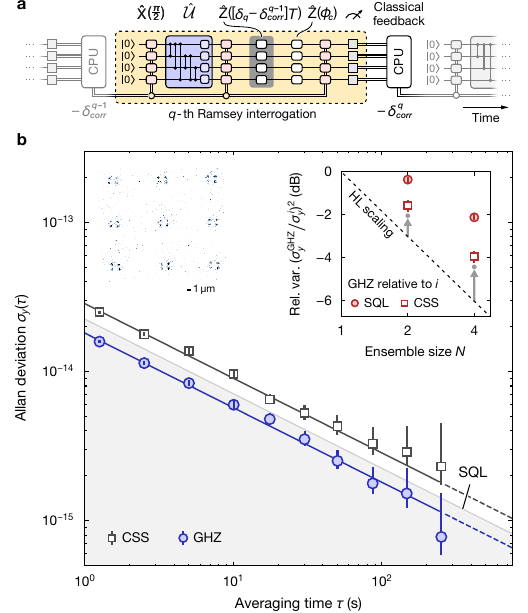}
    \caption{\label{fig3}%
    \textbf{Atom-laser frequency comparisons with GHZ states.}
    \textbf{a},~Equivalent quantum circuit for repeated Ramsey interrogation of the clock laser (atom-laser detuning $\delta_q$) with a GHZ state during the dark time $T$.
    The $q$-th interrogation produces the correction signal~$-\delta_\mathrm{corr}^q$ via classical feedback from a servo running on a computer (CPU).
    This correction signal is applied to the frequency of the clock laser pulses [$\hat{X}(\pi/2)$ rotations] in the $(q+1)$-th interrogation performed in the following experimental cycle.
    \textbf{b},~Overlapping Allan deviation characterizing the fractional frequency instability in an atom-laser comparison for $M=9$ copies of ensembles with size $N=4$ (top left single-shot fluorescence image) and dark time $T=\SI{3}{\milli\second}$.
    It is noted that this quantity is determined from the linear phase estimator at the input of the servo and not from the correction signal $\{-\delta_\mathrm{corr}^q\}$ (see Methods).
    The gray shaded area indicates the region of improved performance with respect to the SQL for $M \times N = 36$ atoms.
    The instability, extracted from the fits shown as solid lines, for the GHZ states (CSS) are $\gainfourrelsql$ below [$\cssfourrelsql$ above] the SQL.
    Due to imperfect qubit initialization (see Methods), the mean total atom number per cycle was $\approx 34$ for both.
    (inset)~By using similar fits, we show how the squared Allan variance at fixed~$\tau$ scales relative to the SQL (red circles) or the CSS (red squares) for variable ensemble size~$N$.
    The arrows pointing to gray circles indicate the theoretically expected variance relative to the SQL after taking into account the reduced parity contrast and fluctuating ensemble size observed in the experiment.
    }
\end{figure}
 
The ratio of sensitivity to QPN of a quantum state is critical in determining the precision of a quantum measurement. The phase sensitivity of an ideal $N$-atom GHZ state is $N$-times enhanced (see Fig.~\ref{fig2}b, up to contrast reduction) compared to a coherent spin state (CSS) of unentangled particles. Because only a single binary parity outcome is obtained from those $N$-atoms, the QPN increases by $\sqrt{N}$. Altogether, this yields the $\sqrt{N}$ improvement in precision that suggests the potential for GHZ states to reach the HL.

More concretely for optical clocks, the basic mode of operation is to synchronize the output of a laser to an atomic reference by regularly inferring the atom-laser detuning from measurements of the atomic populations. The critical metric characterizing the performance of this procedure is the fractional frequency instability~\cite{ludlow2015optical}. Using Ramsey interrogation, the instability for $M$ copies of $N$-atom GHZ states interrogated on each measurement cycle is bounded from below by
\begin{align}
    \sigma_y^{\mathrm{HL}}(\tau) = \frac{1}{2 \pi \nu_0 T \sqrt{M} N} \sqrt{\frac{T_{\rm cycle}}{\tau}}.
\end{align}
Here $\nu_0$ is the clock transition frequency, $T$ is the Ramsey dark time, $T_{\rm cycle}$ the time for a single experimental cycle and $\tau$ the averaging time. For a fixed total atom number per cycle $M \times N$ and all other parameters held constant, the above bound is reduced by $\sqrt{N}$ compared to the SQL achieved by an ideal CSS~\cite{ludlow2015optical}, and can be interpreted as the HL for clock instability in the ensemble size $N$ (but not the total atom number $M \times N$).

To test this paradigm, we investigate the performance of the prepared GHZ states in an atom-laser frequency comparison, employing Ramsey interrogation at a short dark time of $T=\SI{3}{\milli\second}$; $T$ is chosen conservatively to be well within the GHZ-state coherence time, and $T_{\rm cycle} \approx \SI{1.26}{\second}$ for these experiments. The protocol is similar to the coherence time measurements with a fixed readout phase $\phi_c$; additionally, the parity measurement from each experimental cycle is converted into an atom-laser detuning estimate which is used to correct the clock laser frequency on the next cycle (see Fig.~\ref{fig3}a and Methods). Fig.~\ref{fig3}b shows the overlapping Allan deviation which characterizes the atom-laser instability for $M=9$ copies of $N=4$ GHZ ensembles (see top-left image). The GHZ states operate at a fractional frequency instability of $\adevfour$; the Allan variance reduction is $\gainfourrelcss$ compared to the correspondingly prepared CSS and $\gainfourrelsql$ compared to the SQL for $M \times N=36$ total atoms. A summary of the variance reduction for $N=2,4$ is shown in the top-right inset. We observe the improvement growing both with respect to the CSS and the SQL, though the reduction relative to the CSS remains short of the naively expected HL (dashed line). Two contributions to this are parity contrast reduction and averaging over smaller, less-sensitive GHZ states due to imperfect rearrangement (see Methods); correcting the HL scaling for these effects (arrows pointing to gray circles) accounts for most of the discrepancy.

The metrological gain of a single GHZ-state size can be practically harnessed for a restricted class of problems, such as stabilizing certain forms of laser noise~\cite{leroux2017line} or sensing of time-varying signals at a specific bandwidth~\cite{colombo2022entanglement}. However, a key factor in achievable optical clock precision is the atom-laser coherence time~\cite{matei20171,oelker2019demonstration}. For a CSS, this coherence time limit is set by the condition that the integrated Ramsey phase of the stochastically varying atom-laser detuning must have sufficiently high probability to remain within the interval $\left[ -\pi/2, \pi/2 \right]$; this interval is the dynamic range over which the atomic readout can be unambiguously converted into a detuning estimate. Since the parity of an $N$-atom GHZ state oscillates $N$-times more rapidly with phase, the width of this interval is reduced by a factor of $N$. The optimal dark time for the GHZ state is thus $N$-times shorter, cancelling out the increased sensitivity. For the results presented here, we note that the coherence-time limit is set by magnetic-field noise as opposed to laser frequency noise.

\begin{figure*}[t!]
    \centering
    \includegraphics[width=\textwidth]{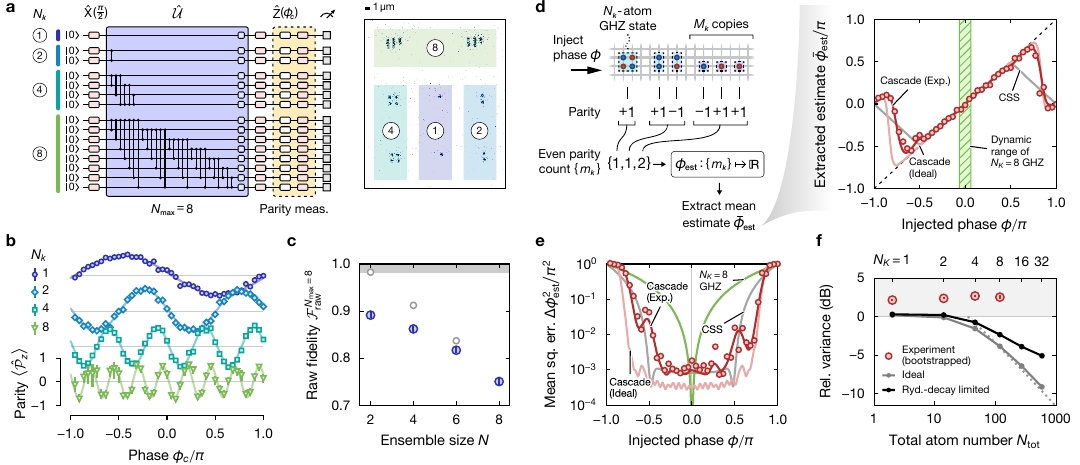}
    \caption{\label{fig4}%
    \textbf{Preparing cascaded GHZ states for multi-ensemble metrology.}
    \textbf{a}, (left) Equivalent quantum circuit for preparing $K=4$ cascaded GHZ states using a global multi-qubit gate ($N_\mathrm{max} = 8$) and $N_k=2^{k-1}$. 
    (right) Single-shot image of atoms rearranged into different ensemble sizes.
    The color shading and circled numbers indicate the group of atoms in the image and quantum circuit.
    \textbf{b}, $N_k$-particle parity (colored markers) and corresponding sinusoidal fits (colored lines, see Methods) for $K=4$ cascaded GHZ states prepared simultaneously within a single experimental shot (see panel~a).
    The different $N_k$ are offset vertically for visual clarity.
    \textbf{c}, Raw GHZ-state fidelity (filled blue circles) for variable ensemble size~$N$ and constant $N_\mathrm{max} = 8$.
    The gray circles correspond to the fidelities for $N_\mathrm{max} = N$ from Fig.~\ref{fig2}.
    The gray shaded region indicates the Rydberg-decay limit for $N=\Nmax=8$.
    \textbf{d}, Phase estimation with cascaded GHZ states.
    (left) A single cascade produces a set of binomial outcomes $\lbrace m_k \rbrace$ characterizing the number of even parity measurements for each size $k$, which is converted to an individual estimate of $\phi$ by an estimator function $\phi_{\rm est} (\lbrace m_k \rbrace)$. 
    (right) Mean phase estimate $\bar{\phi}_{\rm est}$ (red circles) obtained by bootstrapping the cascaded GHZ-state parity measurements in panel~b for $M_K = 2$, $\mu = 8$, and bootstrapped total atom number $N_{\rm tot} = 118$ (see main text). 
    Dark (light) red line shows calculation assuming binomial distributions and fitted (perfect contrast) parity models for each GHZ state. 
    Solid gray line indicates corresponding estimate for a CSS of $N_{\rm tot} = 118$ atoms. 
    The green hatched area indicates the inversion interval for the maximum GHZ-state size $N_K = 8$.
    \textbf{e}, Mean-squared error of the estimator for the same cascade parameters as in panel~d.
    \textbf{f}, Measurement variance relative to SQL $\Delta \phi_{\rm eff}^2 N_{\rm tot}$ (see main text) for varying bootstrapped total atom number $N_{\rm tot}$ and maximum GHZ-state size $N_K$. 
    The experimental results (red circles) are all obtained from the same $K=4$ data.
    The black line is a calculation assuming GHZ-state fidelities limited by Rydberg decay (see Fig.~\ref{fig2}b, $N_{K=6}=32$ extrapolated from scaling in Fig.~\ref{fig1}b), and contrast further reduced by $0.99^{N_k}$ to capture the effect of the currently achieved measurement error rate (see Methods).
    The solid gray line is a calculation assuming perfect contrast for all GHZ-state sizes.
    The dotted line is a reference showing $\pi^2 \ln (N_{\rm tot})/N_{\rm tot}$ (see Methods).
    }
\end{figure*}

\subsection*{Cascaded GHZ-state phase estimation}

Extending the dynamic range is critical for allowing GHZ states to reach HL scaling of clock stability at the optimal dark time. In the entanglement-free context of multi-pass interferometry, a similar hurdle was overcome using protocols resembling the quantum phase estimation algorithm~\cite{higgins2009demonstrating,berry2009perform}; extensions of this scheme to optical clocks with GHZ states were proposed in Refs.~\cite{kessler2014heisenberg,komar2014quantum}. The essential idea is to bridge the gap in dynamic range between the CSS and a large GHZ state by using a cascade of $K$ steadily increasing GHZ-state sizes $N_k$ ($k=1, \hdots, K$, each with $M_k$ copies); each $N_k$ sufficiently updates the prior information on the phase such that the estimate by $N_{k+1}$ is no longer ambiguous. For instance, a phase estimate with $K$ bits of precision could utilize sizes $N_k = 2^{k-1}$ such that the $k$-th ensemble size determines the $k$-th bit of precision. Importantly, near-HL scaling of clock performance is expected to be maintained despite the extra allocation of resources~\cite{kessler2014heisenberg,komar2014quantum}.

To produce cascades, we exploit an important feature of the multi-qubit gate $\UGHZ$: since $\UGHZ$ applies all pairwise CZ gates within an ensemble, regardless of the number of qubits in the ensemble, a single global gate sequence can produce a GHZ state for any $N$ (or specifically in our Rydberg implementation, any $N \leq \Nmax$). This enables the simultaneous generation of multiple GHZ-state sizes without additional local controls beyond initialization of the qubit ensembles (see Fig.~\ref{fig4}a). In Fig.~\ref{fig4}b, we demonstrate the preparation and parity readout of a GHZ-state cascade with $K=4$ and $N_k=2^{k-1}$ using the multi-qubit gate for $\Nmax = 8$. For these data, we attempt to prepare $M_k = 2$ copies of each size on each experimental cycle (see Fig.~\ref{fig4}a). While this scheme benefits from reduced complexity, it suffers from degraded parity contrast of ensembles $N < \Nmax$, as shown in Fig.~\ref{fig4}c. 

In Fig.~\ref{fig4}d, we explore phase estimation with cascaded GHZ states to demonstrate their extended dynamic range. The zero dark time data from Fig.~\ref{fig4}b is reanalyzed to interpret the analysis phase $\phi_c$ as an unknown parameter $\phi$ which we would like to determine from the parity measurement. An appropriate estimator function is used to convert a set of parity outcomes obtained from a single cascade measurement into a phase estimate $\phi_{\rm est}$ (see Fig.~\ref{fig4}d and Methods). The estimator we use is optimized for a Gaussian prior, which models the laser phase diffusion typically encountered in atomic clock operation; a standard deviation $\sigma_{\phi} = \pi/6$ is chosen to be larger than the inversion range of the maximum-size GHZ state. Repeating the measurement many times yields the mean estimate $\bar{\phi}_{\rm est}$ and the mean-squared error (MSE) $\Delta \phi_{\rm est}^2$. 

In the right panel of Fig.~\ref{fig4}d, $\bar{\phi}_{\rm est}$ from a $K=4$ cascade with a linear distribution of copies $M_k = M_K + \mu (K-k)$ for $M_K=2$ and $\mu=8$ (bootstrapped total atom number $N_{\rm tot} = 118$) is shown, revealing that unbiased estimation is recovered over a large fraction of the $2\pi$-interval. Due to limitations in the number of ensembles that we can prepare simultaneously, this data is obtained by bootstrap resampling over all repeated measurements at a single $\phi$ (see Methods). We do this to investigate cascades with more copies at smaller ensemble sizes, which helps to mitigate large estimation errors~\cite{higgins2009demonstrating,berry2009perform}. The experimental cascade (dark red) has only a slightly larger MSE than that of a near-unity contrast CSS (gray) with the same $N_{\rm tot}$. A cascade with perfect parity contrast (light red) would have significantly reduced MSE over almost the entire range. In contrast, the MSE for multiple copies of just the largest $N_K=8$ GHZ state (green) is small only in a narrow region about $\phi=0$. 

While the MSE is measured at zero dark time $T=0$, we are primarily interested in the performance of the cascade during clock operation with $T > 0$. The effective measurement uncertainty $\Delta \phi_{\rm eff}$ associated with a cycle of Ramsey interrogation can be inferred by incorporating a prior which reflects the distribution in integrated atom-laser detuning at a specific $T$~\cite{leroux2017line,kaubruegger2021quantum,marciniak2022optimal} (see Methods). Using the same $\sigma_{\phi} = \pi/6$ Gaussian prior as used for the phase estimator, we find that the current experimental results, all computed from the same $K=4$ cascade data, are only 2~dB above the SQL in effective measurement variance. A major current limitation is the reduction of contrast for smaller ensembles being subjected to a larger $\Nmax$ gate (see Fig.~\ref{fig4}c). Looking towards the future, we compute the variance reduction up to $K=6$ assuming fidelities limited by Rydberg decay and measurement error (see Fig.~\ref{fig4}f caption). With this realistic model, the cascade is expected to demonstrate a substantial improvement for hundreds of atoms (black). Without measurement errors, the variance reduction follows closely to that of a perfect contrast cascade (gray); the scaling is empirically found to be near the HL with both constant and logarithmic overheads (see Methods).

\subsection*{Conclusion}
We have demonstrated high-fidelity two-qubit entangling gates and used multi-qubit gates to prepare GHZ states of up to 9 optical clock qubits. Employing these GHZ states for metrology, we have performed an atom-laser frequency comparison below the SQL and extended the phase estimation dynamic range with a multi-ensemble GHZ-state cascade; the latter capability restores the compatibility of large GHZ states with the long dark times available for unentangled atoms when local oscillator noise dominates, as is the case for the state-of-the-art optical lattice clocks.
These results establish key building blocks for GHZ-based optical clocks operating near the HL~\cite{kessler2014heisenberg}, which may also serve as a critical element for remotely-entangled clock networks~\cite{komar2014quantum, nichol2022elementary}. Near-term goals involve further improving Rydberg gate fidelities while combining these high-fidelity clock-qubit controls with recent advances in scaling to larger atom arrays~\cite{norcia2024iterative,gyger2024continuous}. A current limitation to cascade performance is contrast reduction for smaller ensembles; this issue could be mitigated by shelving coherence in other degrees of freedom~\cite{lis2023midcircuit, scholl2023erasure2} or using coherence-preserving moves~\cite{shaw2024multi, finkelstein2024universal} with entangling zones~\cite{bluvstein2022quantum}. 

Besides GHZ states, the high-fidelity entangling operations demonstrated also pair well with complementary strategies for generating metrological enhancements, such as hardware-oriented variational optimization~\cite{kaubruegger2019variational, kaubruegger2021quantum, marciniak2022optimal}. Comparing different entanglement strategies, ranging from spin-squeezing~\cite{eckner2023realizing} to GHZ-state generation, on their practical utility, accounting for trade-offs in metrological gain and robustness, is an interesting avenue for programmable clocks. Beyond metrology, the multi-qubit gate technique demonstrated here can be extended to any diagonal, symmetric phase gate in principle, such as the multi-qubit controlled-Z gate~\cite{evered2023high}. 


%

\clearpage

\section*{Methods}

\subsection*{State detection}
To determine the population in the computational states, we employ a detection scheme to map $\ket{0}$ ($\ket{1}$) to being dark (bright) in a fluorescence image. Our detection scheme begins by employing a push-out pulse using resonant $\SI{461}{\nano\meter}$ light to remove atoms in $\ket{0}$. These atoms are successfully removed with probability $0.9999(1)$. We then apply a clock $\pi$-pulse to transfer $\ket{1} \to \ket{0}$; this mitigates Raman scattering of the clock state (see Effective state decay section) during the $\approx \SI{30}{\milli \second}$ period over which we ramp off a large magnetic field in preparation for imaging. At the low-field condition, we additionally apply $\SI{679}{\nano \meter}$ and $\SI{707}{\nano \meter}$ repumping light which is intended to drive any remaining population in $\ket{1}$ back to $\ket{0}$; note, however, that any inadvertent population in \tripptwo will also be repumped. 

The atoms are then imaged by driving the ground $\singszero \leftrightarrow \singpone$ transition while simultaneously sideband cooling on the $\singszero \leftrightarrow \trippone$ transition.
For most data in the main text, we use a long exposure time of $\SI{300}{\milli\second}$.
For the data in Fig.~\ref{fig2}d and Fig.~\ref{fig3}b (as well as most of the Extended Data Figures), we use a shorter exposure time of $\SI{100}{\milli\second}$ to increase the data acquisition rate at the cost of slightly increased imaging infidelity.

We estimate the imaging infidelity and loss by characterizing the disagreement of two subsequently taken fluorescence images of the same atomic sample, but taken with different exposure times.
Here, the first image has a much longer exposure time of $\SI{1200}{\milli\second}$ to significantly lower the imaging infidelity (estimated from the photon count histogram).
This allows us to treat this image as the ground truth after correcting for imaging loss which we determine independently.
By comparing the measurement result of this first image, i.e., whether a site is identified as bright or dark, to the second $\SI{300}{\milli\second}$-long image, we obtain an estimate for the imaging infidelity.
For the rearrangement pattern corresponding to the Bell-state measurements, the inferred probabilities of identifying a dark site incorrectly as bright or a bright site incorrectly as dark typically take values $p_{d \rightarrow b} \approx 0.002$ and $p_{b \rightarrow d} \approx 0.002$, respectively.
We note that these probabilities are significantly increased up to $p_{d \rightarrow b} \approx 0.009$ and $p_{b \rightarrow d} \approx 0.003$ for the larger ensemble sizes where the atoms are rearranged into patterns with a single lattice site spacing along one direction (see Extended Data Table~\ref{tabed1}).
For reported measurement-corrected fidelities of Bell states ($N=2$) and $N$-atom GHZ states (see Extended Data Table~\ref{tabed2}), we account for the imaging infidelity determined for representative rearrangement patterns.

\subsection*{Rydberg excitation}
Our Rydberg laser system has been described in detail before~\cite{schine2022long}, though some modifications have been made for this work. Here we mainly describe aspects related to pulse generation for Rydberg gates. $\SI{317}{\nano\meter}$ ultraviolet (UV) light is sent through an acousto-optic modulator (AOM) (AA Opto-Electronics MQ240-A0,2-UV) in single-pass configuration to control the beam's phase and intensity. We measure a rise time of $\approx \SI{15}{\nano\second}$. The radio frequency (RF) tone for driving the AOM is generated by an arbitrary waveform generator (AWG) built in-house by the JILA electronics shop. The Rydberg laser is phase modulated by programming the AWG output phase, which can be updated in $\SI{6.5}{\nano\second}$ steps. To clean up the spatial mode and suppress pointing fluctuations, the first-order diffracted beam through the AOM is sent through a short ($\leq \SI{1.5}{\meter}$) hydrogen-loaded, UV-cured photonic crystal fiber~\cite{colombe2014single} before being focused down on the atoms. 

A small fraction of the fiber output is diverted to a photodetector (Thorlabs APD130A2) which is used to perform a sample and hold of the UV intensity for mitigation of shot-to-shot Rabi frequency fluctuations; we measure a fractional standard deviation in integrated pulse area of 0.007-0.008. A limitation in the current setup is conversion of phase modulation to intensity modulation; the phase modulation alters the instantaneous RF frequency and thus the deflection angle of the AOM diffraction, which then leads to variable fiber coupling efficiency. We mitigate this effect by careful alignment to the fiber, but residual modulation at the 5-10\% level was observed for the larger $\Nmax$ gates. We also perform ex-situ heterodyne measurements of the first and zero-order modes of the AOM before the fiber to benchmark the transduction of RF phase to optical phase; these measurements use a higher bandwidth photodetector (Thorlabs APD430A2) compared to the one used for the sample and hold. We did not observe any significant distortions, and thus did not apply any corrections to the numerically optimized waveforms programmed into the AWG. We note that there may be phase distortions introduced by the fiber which we did not test~\cite{ma2023high}.

\subsection*{Lattice release and recapture}
We turn off the lattice during Rydberg excitation to eliminate anti-trapping effects and spatially inhomogeneous Stark shifts. However, this causes heating and imperfect recapture of the atoms. The combination of single-photon Rydberg excitation and the optical lattice used in this work causes this to be a significant effect, particularly for the longer multi-qubit gates. Adiabatically ramping the lattice to lower depths before release helps to alleviate this issue, but the depth cannot be set arbitrarily low due to tunneling. Based on gate fidelities, we empirically found that ramping to a lattice depth of around 50$\,E_r$ before quenching off was optimal; $E_r = \hbar^2 k_l^2/2m \approx 2 \pi \hbar \times \SI{3.4}{\kilo \hertz}$, with $\hbar$ the reduced Planck constant, is the single-photon recoil energy of the $\lambda = 2 \pi/k_l = \SI{813.4275}{\nano\meter}$ photons used to generate our 2D bowtie lattice~\cite{schine2022long,young2022tweezer}.

To more quantitatively understand the magnitude of errors, we develop a model for release and recapture. We treat this as free-expansion of the ground-band Wannier state in the 2D lattice; we ignore expansion along the weakly confined axial direction, but in principle the calculation 
straightforwardly generalizes to 3D. Let $\wanniernR$ denote the Wannier state in the $n$-th band at site $\vec{R}$ and $\blochnq$ denote the Bloch state in the $n$-th band at quasimomenta $\vec{q}$ (the use of $n$ to denote band index is restricted to this section). We will work in units of the bowtie lattice spacing $\lambda/\sqrt{2}$, wavevector $\sqrt{2} k_l$ and energy $2E_r$. Then $\vec{R}$ is a 2D vector of integers denoting sites of the lattice and $\vec{q} = q_x \hat{x} + q_y \hat{y}$ with $q_{x,y} \in \left[-1/2,1/2\right]$ which defines the first Brillouin zone (BZ). The atomic wavefunction after a free-expansion time $t$ [in units of $\hbar/(2E_r)$] is given by $\ket{\psi(t)} = e^{-i \hat{H}_{\rm free} t} \ket{w_{0,0}}$ where $\hat{H}_{\rm free}$ is the kinetic energy Hamiltonian. Using the definition $\ket{w_{n,\vec{R}}} = \int_{\rm BZ} \dd \vec{q} e^{-i \vec{q} \cdot \vec{R}} \ket{\psi_{n,\vec{q}}}$ and the expansion of Bloch states in the plane-wave basis $\ket{\psi_{n,\vec{q}}} = \sum_{\vec{m}} c_{\vec{m}}^{n, \vec{q}} \ket{\vec{q} + \vec{m}}$ ($\vec{m}$ is also a 2D vector of integers), we compute the state overlap with any given Wannier state over time to be 
\begin{align}
    \bra{w_{n, \vec{R}}} \ket{\psi(t)} = \int_{\rm BZ} \dd \vec{q} e^{i \vec{q} \cdot \vec{R}} \sum_{\vec{m}} \left( c_{\vec{m}}^{n, \vec{q}} \right)^* c_{\vec{m}}^{0,\vec{q}} e^{- i (\vec{q} + \vec{m})^2 t}.
    \label{eqwannieroverlap}
\end{align}
The asterisk denotes complex conjugation. The expansion coefficients can be obtained through a band structure calculation, and here are defined such that $\sum_{\vec{m}} \left( c_{\vec{m}}^{n', \vec{q}} \right)^* c_{\vec{m}}^{n,\vec{q}} = \delta_{n',n}$ with $\delta_{n',n}$ the Kronecker delta. Once computed, these Wannier state overlaps can be straightforwardly used to calculate various observables of interest in the experiment.

Here we specifically consider the recapture probability of atoms onto the same site, and the heating of those recaptured atoms. Let $n_{\rm max}$ denote the highest band which remains trapped by the lattice, explicitly determined as the highest band with average energy below the lattice potential maximum. The recapture probability is then
\begin{align}
    p_{\rm recapture} = \sum_{n=0}^{n_{\rm max}} \abs{ \bra{w_{n, 0}} \ket{\psi(t)} }^2.
\end{align}
The heating is characterized by the average phonon number $n_r$ (this notation is also restricted to this section and Extended Data Fig.~\ref{fig3ed}b) of the recaptured atoms
\begin{align}
    \bar{n}_r = \frac{\sum_{n=0}^{n_{\rm max}} n_r \abs{ \bra{w_{n, 0}} \ket{\psi(t)} }^2}{p_{\rm recapture}}.
\end{align}
Note that the band index $n$ differs from the motional quantum number $n_r$ by a combinatorial factor. In $d$-dimensions, there are ${n_r + d -1 \choose d-1 }$ bands with the same $n_r$. Here we consider $d=2$ such that this number is $n_r+1$. While we consider an initial ground-band Wannier state to good approximation for our system, a thermal average over initially occupied higher bands $n_0$ can be performed by replacing $c_{\vec{m}}^{0,\vec{q}} \to c_{\vec{m}}^{n_0,\vec{q}}$ in Eq.~(\ref{eqwannieroverlap}). Finally, the effect of the UV photon recoil is included by modifying the kinetic energy $\left( \vec{q} + \vec{m} \right)^2 \to \left( \vec{q} + \vec{m} + \frac{\vec{k}_{\rm UV}}{\sqrt{2} k_l} \right)^2$ in the same equation; here $\abs{k_{\rm UV}} = 2\pi/\lambda_{\rm UV}$ with $\lambda_{\rm UV} = \SI{317}{\nano\meter}$, and we take $\vec{k}_{\rm UV}$ along the $x$-direction.

In Extended Data Fig.~\ref{fig3ed}b, we perform measurements of survival as a function of trap turn-off duration for both ground and Rydberg state atoms. We observe good agreement of the data with the theory for $p_{\rm recapture}$ developed above. The Rydberg-state data includes Rydberg $\pi$-pulses just after the release and just before the recapture; we suspect infidelity in these pulses accounts for the reduced survival at short times. We fit the quadratic decay at short times to have a Gaussian $1/e$ time constant of $\SI{8.7(1)}{\micro \second}$, though we note that the decay is not Gaussian at later times. For the longest $\Nmax=10$ gate, we expect a recapture loss of $<0.01$ based on this. We also compute $\bar{n}_r$ to get a sense of the degree of heating this effect causes. For all GHZ-state data in the main text, the lattice turn-off duration was $<\SI{2}{\micro \second}$, which suggests $\bar{n}_r$ at the $0.1$ level.

\subsection*{Effective state decay}
Various processes can cause an effective decay of population in $\ket{1}$ and $\ket{r}$ over time. Such processes not only degrade the true GHZ-state fidelity, but also cause leakage out of the computational basis which results in misidentification for our state detection scheme. Thus, it is critical to characterize the degree to which such decay happens.

The natural lifetime of $\ket{1}$ is generally much larger than the relevant time-scales explored in this work. However, atoms in the $^3$P$_J$ manifold undergo Raman scattering in the lattice~\cite{dorshcer2018lattice}. Ultimately, this scattering will depopulate $\ket{1}$ and repopulate $\ket{0}$ due to the much shorter natural lifetime of the \trippone state (see Extended Data Fig.~\ref{fig2ed}a). Let $p_0$, $p_1$ and $p_2$ denote the ground, clock and \tripptwo state populations respectively (this notation is restricted to this section). We model the dynamics of these populations as
\begin{align}
    \dv{p_0}{t} &= \Gamma_{1 \to 0} p_1 + \Gamma_{2 \to 0} p_2, \nonumber  \\
    \dv{p_1}{t} &= - \left( \Gamma_{1 \to 0} + \Gamma_{1 \to 2} \right) p_1, \\
    \dv{p_2}{t} &= \Gamma_{1 \to 2} p_1 - \Gamma_{2 \to 0} p_2. \nonumber 
\end{align}
We experimentally extract the scattering rates $\Gamma$ by fitting the above rate model to measurements of $p_0$ and $p_1$ over time after initializing all atoms in $\ket{1}$, shown in Extended Data Fig.~\ref{fig2ed}c. The fit also includes a separate measurement of $p_1+p_2$ which is not shown; within the rate model, this sum is equivalent to $1-p_0$. In principle, there is a process driving $\tripptwo \to \ket{1}$, but the fit procedure yields a value consistent with zero when this term is included. We obtain scattering rates of $\Gamma_{1 \to 0} = \SI{0.48(1)}{\hertz}$, $\Gamma_{1 \to 2} = \SI{0.26(2)}{\hertz}$, and $\Gamma_{2\to 0} = \SI{0.47(3)}{\hertz}$ for measurements performed at a lattice depth of $\approx 920E_r$; for the rates where calculations have been reported, the fitted results are in good agreement with expectation~\cite{dorshcer2018lattice}. From this we estimate that the decay of initially prepared $\ket{1}$ state is $<0.002$ for most experiments presented, with roughly 1/3 of that population ending up in \tripptwo before the fluorescence image.

Next we characterize the lifetime of the Rydberg state $\ket{r}$. Because there are many paths with which the Rydberg decay may proceed, we follow the protocol discussed in Ref.~\cite{scholl2023erasure} to group the decay into states dark and bright to our detection protocol (see Extended Data Fig.~\ref{fig2ed}a). In both cases, the measurements proceed by initializing all atoms in $\ket{r}$ and waiting a variable duration. To measure dark state decay, we apply a final Rydberg $\pi$-pulse; to measure bright state decay, we apply a final Rydberg auto-ionization pulse~\cite{madjarov2020high}. The survival over time is plotted for these two protocols in Extended Data Fig.~\ref{fig2ed}b. These experiments were performed in optical tweezers and at a fixed trap turn-off duration of $\SI{40}{\micro \second}$ to mitigate the effect of release and recapture; nevertheless, recapture failure accounts for a majority of the population reduction at zero time. We fit the curves simultaneously to the 3-parameter forms $Ae^{-t/\tau_r^{\rm d}}$ and $\left(A \tau_r^{\rm d}/\tau_r^{\rm b} \right)\left( 1 - e^{-t/\tau_r^{\rm b}} \right)$. This yields a $\tau_r^{\rm d} = \SI{51(3)}{\micro\second}$ dark-state and $\tau_r^{\rm b} = \SI{86(3)}{\micro\second}$ bright-state decay time. The expected Rydberg decay contribution is $<0.03$ for the largest $\Nmax=10$ gate used in this work.

\subsection*{Clock and Rydberg coherence}
In order to achieve appreciable Rabi frequency on the $\ket{0} \leftrightarrow \ket{1}$ transition, all experiments in the main text are performed at a magnetic field of $\SI{275}{G}$~\cite{taichenachev2006magnetic}. The clock and Rydberg transition frequencies acquire a substantial sensitivity to field variations at this large bias field due to quadratic Zeeman and diamagnetic shifts respectively. In particular for the clock transition, field fluctuations are the limiting factor in the $\SI{327(1)}{\milli\second}$ CSS atom-laser $1/e$ coherence time shown in Fig.~\ref{fig2}d. One major source of field noise found in the system during this work was a $\SI{0.5}{G}$ peak-to-peak oscillation synchronized with the $\SI{60}{\hertz}$ mains power. To mitigate this effect, we apply a feed-forward to the clock laser frequency to compensate the change in magnetic field. The feed-forward was calibrated by performing clock Rabi spectroscopy as a function of wait time with respect to a specific mains phase. For the Rydberg, where pulses are essentially instantaneous with respect to these mains variations, we rely on performing the pulses at a specific point in the mains phase where the field variation is minimal. In the future, active stabilization of the magnetic field will be used to mitigate this effect.

Another significant contribution to coherence reduction is non-zero temperature. In Extended Data Fig.~\ref{fig2ed}c, we show clock Rabi oscillations over many coherent cycles. We believe the contrast reduction at later times arises from imperfect motional state cooling, and we fit the data to obtain a 1D ground state fraction of 0.96(1) along the clock-laser propagation direction; we note that in contrast to the direction shown in Fig.~\ref{fig1}a (which was chosen for visual clarity), the clock laser actually propagates at a significant angle relative to the 2D lattice axes (but still in the plane). This Rabi-oscillation measurement does not include heating due to the release and recapture.

We also perform Rydberg Rabi and Ramsey dephasing measurements, shown in Extended Data Fig.~\ref{fig2ed}b. In both cases, we fix the total lattice turn-off duration to $\SI{5}{\micro\second}$, independent of the Rabi/Ramsey time. These data are used to estimate an upper bound on inhomogeneous fluctuations in the Rabi frequency $\Omega_r$ and detuning $\Delta$, which we assume to be characterized by Gaussians with standard deviation $\sigma_{\Omega}$ and $\sigma_{\Delta}$. Fitting to a Monte Carlo simulation, we find a fractional Rabi frequency standard deviation of $\sigma_{\Omega}/\Omega_r = 0.0055(7)$ and a detuning standard deviation of $\sigma_{\Delta}/(2\pi) = \SI{49(2)}{\kilo\hertz}$. The fractional Rabi frequency fluctuations are slightly larger than would be expected based on pulse area fluctuations as monitored on a photodetector, which could be attributed to pointing fluctuations or spatial inhomogeneity of the Rydberg laser. For the Ramsey dephasing, we estimate that Doppler dephasing yields a contribution of $\SI{27}{\kilo\hertz}$ standard deviation; the remainder we expect arises from a combination of magnetic field, electric field and laser phase noise.

\subsection*{Clock and Rydberg rotation fidelity}

High fidelity clock and Rydberg rotations are crucial to generating clock-qubit GHZ states. We characterize our Rydberg and clock $\pi$-pulse fidelities in Extended Data Fig.~\ref{fig2ed}d and e. Fidelities are extracted by fitting to a parabolic form. For the clock, we find a raw $\pi$-pulse fidelity of $\Fclock$. A majority of the error is accounted for by imaging loss and infidelity and lattice Raman scattering. These data were performed at $920E_r$, with typical depths for clock operations ranging from 830--920$E_r$. For the Rydberg, we characterize both the single-atom and blockaded two-atom $\pi$-pulses. An auto-ionization pulse, with an auto-ionization timescale of $\SI{0.32(1)}{\micro\second}$, was used to achieve a Rydberg state detection fidelity of $0.995(1)$. The data shown are corrected for state preparation and measurement (SPAM) errors following the procedure described in Ref.~\cite{madjarov2020high}; the correction includes imaging loss and infidelity, clock state transfer fidelity, and Rydberg state detection fidelity. The SPAM-corrected fidelities are $\Frydspam$ for single atoms and $\Frydbspam$ for pairs of blockaded atoms.

\subsection*{GHZ preparation and fidelity measurement}
In this work, GHZ states are prepared using a combination of global single-qubit clock rotations $\hat{X}(\theta), \hat{Z}(\theta)$ and the multi-qubit gate $\UGHZ$. Here, $\theta$ denotes the angle of rotation. Explicitly, for $\hat{X}$ rotations on $N$-atoms we have $\hat{X}(\theta) = \prod_{j=1}^N \exp \left(-i \frac{\theta}{2} \hat{\sigma}_x^j \right)$, where $\hat{\sigma}_x^j$ is the $x$ Pauli operator acting on the $j$-th atom; an analogous form exists for $Z$ rotations with $\hat{\sigma}_x^j \to \hat{\sigma}_z^j$. Starting with the product state $\ket{0}^{\otimes N}$, we apply $\hat{X}(\pi/2) \hat{Z}(\alpha_c) \UGHZ \hat{X}(\pi/2)$ to produce the GHZ state. While the exact form of $\UGHZ$ requires $\alpha_c = 0$ (see fully connected graph state from $\UGHZ$), the Rydberg implementation causes an additional single-particle phase. We experimentally calibrate $\alpha_c$ by scanning the clock laser phase before the final $\hat{X}(\pi/2)$ gate and maximizing the observed GHZ populations $p_0+p_N$. To generate the Bell state, we instead applied the circuit $\hat{X}(-\pi/4) \hat{Z}(\alpha_c) \hat{\mathcal{U}}_{\rm CZ} \hat{X}(\pi/2)$ with $\hat{\mathcal{U}}_{\rm CZ} = e^{i \pi \hat{n}_1 \hat{n}_2}$ the CZ gate.

For an experimentally prepared density matrix $\hat{\rho}$, the GHZ-state fidelity can be defined as $\mathcal{F} = \mathrm{max}_{\theta} \left[ \bra{\mathrm{GHZ}} \hat{Z}(-\theta) \hat{\rho} \hat{Z}(\theta) \ket{\mathrm{GHZ}} \right]$. We characterize $\mathcal{F}$ by measuring the populations in $\ket{0}^{\otimes N}$ and $\ket{1}^{\otimes N}$, along with the coherence between those states. We obtain the populations by repeated measurements of $p_0 + p_N$ at the calibrated value of $\alpha_c$; $p_0$ ($p_N$) describes the probability of measuring $n=0$ ($n=N$) atoms in $\ket{1}$. We obtain the coherence by taking parity measurements after applying additional single-qubit analysis rotations $\hat{X}(\pi/2)\hat{Z}(\phi_c)$ with variable angle $\phi_c$. For our measurement basis, the parity operator is given by $\parityz = (-1)^N e^{i \pi \hat{n}} = \prod_{j=1}^N \hat{\sigma}_z^j$. The $N$-atom GHZ-state coherence is extracted from fitting the oscillation of the parity expectation to the form $C \sin \left[ N (\phi_c - \phi_0) \right]+y_0$; $C$ is the contrast characterizing the coherence, and $\phi_0$ and $y_0$ are additional fitting parameters.

\subsection*{Fully connected graph state from $\UGHZ$}

A graph state is associated with a graph $G = (V,E)$ consisting of a set of vertices $V$ (representing qubits) which are connected by a set of edges $E$ (representing CZ gates). Starting from the product state $\ket{+_x}^{\otimes V}$ where $\ket{+_x} = \left(\ket{0} + \ket{1} \right)/\sqrt{2}$, the graph state $\ket{G}$ can be defined up to a global phase by~\cite{hein2004multiparty}
\begin{align}
    \ket{G} = \prod_{(a,b) \in E} \hat{\mathcal{U}}_{\rm CZ}^{(a,b)} \ket{+_x}^{\otimes V}.
\end{align}
Here $\hat{\mathcal{U}}_{\rm CZ}^{(a,b)} = e^{i \pi \hat{n}_a \hat{n}_b}$ is a CZ gate acting on the qubits at the vertices $a,b \in V$, or equivalently the qubits connected by the edge $(a,b) \in E$.

The form $\UGHZ$, given in Eq.~\eqref{eq:UGHZ}, can be understood by expanding out $\hat{n}^2 = \sum_{j=1}^N \hat{n}_j^2 + \sum_{j < k} 2 \hat{n}_j \hat{n}_k$. Noting that $\hat{n}_j^2 = \hat{n}_j$, $\UGHZ$ can be re-expressed as 
\begin{align}
    \UGHZ = e^{i N \pi/4} \hat{Z} \left( -\frac{\pi}{2} \right) \exp \left( i \pi \sum_{j < k} \hat{n}_j \hat{n}_k \right).
\end{align}
The third factor describes performing a CZ gate on each pair of qubits. By applying this to $\ket{+_y}^{\otimes N} = \hat{X}(\pi/2) \ket{0}^{\otimes N}$ with $\ket{+_y} = \left( \ket{0} + i \ket{1} \right)/\sqrt{2}$, we obtain the graph state $\ket{G}$ (up to a global phase) associated with the fully connected graph $G$ of $N$-vertices, in which there is an edge between all vertex pairs. The fully connected graph is equivalent to the GHZ state under local unitary operations~\cite{hein2004multiparty}.

Here we explicitly show that $\hat{X}(\pi/2) \UGHZ \hat{X}(\pi/2)$ produces the GHZ state. We begin by noting that $\mathcal{U} = 1$ $(i)$ for even (odd) $n$. It is straightforward to see then that $\UGHZ$ can be expressed as 
\begin{align}
    \UGHZ = \frac{1+i}{2} \hat{I} + \frac{1-i}{2} (-1)^N \parityz.
\end{align}
Here $\hat{I}$ denotes the identity. Noting that $\hat{X}(\pi/2) \hat{\sigma}_z^j \hat{X}(\pi/2) = \hat{\sigma}_z^j$ and $\hat{X}(\pi) = -i \hat{\sigma}_x^j$, we then have 
\begin{align}
    \hat{X} \left( \frac{\pi}{2} \right) \UGHZ \hat{X} \left( \frac{\pi}{2} \right) = \frac{e^{-i \frac{\pi}{4}}}{\sqrt{2}} \left[ (-i)^{N-1} \hat{\mathcal{P}}_x + (-1)^N \parityz \right],
\end{align}
where $\hat{\mathcal{P}}_x = \prod_{j=1}^N \hat{\sigma}_x^j$, which is the parity along a different axis. Applying this to $\ket{0}^{\otimes N}$, we obtain the GHZ state
\begin{align}
    \hat{X} \left(\frac{\pi}{2} \right) \UGHZ \hat{X} \left( \frac{\pi}{2} \right) \ket{0}^{\otimes N} = \frac{e^{-i \frac{\pi}{4}}}{\sqrt{2}} \left( \ket{0}^{\otimes N} + (-i)^{N-1} \ket{1}^{\otimes N} \right).
\end{align}
Applying an additional global $\hat{Z} \left[ - \frac{(N-1) \pi}{2 N} \right]$ rotation yields the form $\ket{\mathrm{GHZ}}$ in Eq.~\eqref{eqghz} up to a global phase.

\subsection*{Optimal control for multi-qubit gates}

To find optimal Rydberg pulses for implementing $\UGHZ$, we closely follow the protocol described in Ref.~\cite{jandura2022time}. We consider a time-dependent Rydberg coupling of the form $\Omega_r e^{-i \phi_r(t)}$. We assume an infinite Rydberg blockade strength such that the dynamics of each excitation sector $n$ can be described by considering an arbitrary product state $\ket{\psi_n} = \ket{0}^{\otimes(N-n)} \ket{1}^{\otimes n}$ and a corresponding W-state $\ket{W_n}$ of a single Rydberg excitation~\cite{lukin2001dipole, dudin2012observation, zeiher2015microscopic, bernien2017probing}. $\ket{\psi_n}$ is evolved for duration $\Tgate$ under the two-level Hamiltonian 
\begin{align}
    \hat{H}_n = \frac{\sqrt{n} \Omega_r}{2} \left[ \cos \phi_r(t) \hat{\sigma}_{x,r} + \sin \phi_r(t) \hat{\sigma}_{y,r} \right] - i \frac{\gamma_r}{2} \ket{W_n} \bra{W_n}.
\end{align} 
$\hat{\sigma}_{x(y),r}$ denote the Pauli operators on the two-level subspace spanned by $\ket{\psi_n}$ and $\ket{W_n}$. We include a non-Hermitian loss at rate $\gamma_r = \gamma_r^{\rm d} + \gamma_r^{\rm b}$ (see Extended Data Fig.~\ref{fig2ed} and effective state decay section) to estimate optimal achievable fidelities given accessible Rydberg parameters. Additionally, we multiply $\Omega_r$ by a time-dependent envelope function to capture finite rise-time effects on the experiment. The figure of merit to optimize is explicitly given by
\begin{align}
    F = \frac{1}{4^N} \mathrm{max}_{\alpha_c} \left[ \abs{\sum_{n=0}^N \begin{pmatrix} N \\ n \end{pmatrix} i^{n^2} e^{i n \alpha_c} \bra{\psi_n} \ket{\psi_n(\Tgate)}}^2 \right].
\end{align}
A discretized form for $\phi_r(t)$ is assumed to utilize GRAPE~\cite{khane2005optimal}; the time-step is naturally set by the update rate for the AWG performing the modulation. We use a first-order approximation for the gradient of $F$ with respect to the control phase, and employ the Broyden–Fletcher–Goldfarb–Shanno algorithm for gradient descent.

\subsection*{Ensemble size scaling for multi-qubit Rydberg gates}

The largest ensemble size the multi-qubit gate can be successfully applied to depends on the number of atoms $N_b$ that can be placed in a single Rydberg blockade radius $R_b$. Here we outline general considerations for how $N_b$ scales with Rydberg principal quantum number $n$ (notation restricted to this section). For these arguments, we assume a fixed laser intensity; other conditions can be reasonably considered, such as fixed Rabi frequency or decay fraction, but do not change the qualitative conclusions.

The Rydberg blockade radius is given by $R_{\rm b} = \left( C_6 / \Omega_r  \right)^{1/6}$. The $C_6$ interaction coefficient and Rydberg Rabi frequency $\Omega_r$ vary as $C_6 \propto n^{11}$ and $\Omega_r \propto n^{-3/2}$~\cite{low2012experimental}, yielding a scaling of the blockade radius $R_{\rm b} \propto n^{25/12}$. Because atoms cannot be placed arbitrarily close together, $N_b$ is additionally limited by a minimum spacing $R_{\rm min}$. In two dimensions, the number of atoms fitting in the blockade radius then scales as $N_{\rm b} \propto \left( R_{\rm b}/R_{\rm min} \right)^2$. Experimentally, $R_{\rm min}$ could be set by the lattice spacing $a_{\rm lat}$ in optical lattices, or alternatively the beam waist for optical tweezers; this spacing is independent of $n$, yielding a scaling $N_{\rm b} \propto n^{25/6}$ favoring larger $n$. However, a separate limitation for $R_{\rm min}$ is the presence of molecular resonances at small interatomic spacings which can drastically degrade the blockading interaction at certain separations. To avoid these effects, one can restrict to placing atoms outside the radius $R_{\times}$ of the outermost resonance, which can be estimated to scale as $R_{\times} \propto n^{8/3}$~\cite{derevianko2015effects}; this then yields $N_{\rm b} \propto n^{-7/6}$ which instead favors lower $n$.

In practice, the $n^{25/6}$ scaling will apply for small $n$ where $R_{\times} \ll a_{\rm lat}$, and the $n^{-7/6}$ scaling should apply for large $n$ where $R_{\times} \gg a_{\rm lat}$. From this, we generally expect that the maximum $N_b$ will be achieved for $n$ such that $R_{\times} \sim a_{\rm lat}$. Because the impact of molecular resonances varies drastically with interatomic spacing, the limitation imposed by $R_{\times}$ may be partially circumvented by fine-tuning the atomic separation~\cite{derevianko2015effects}; this requires accurate modeling for the Rydberg series of interest and careful atomic positioning, and the suppression improves for tighter atomic confinements. While we did not intentionally engineer such a suppression for this work, we note that such an effect may be relevant for the $N\geq6$ data with single lattice site spacing as $a_{\rm lat} < R_{\times} < 2 a_{\rm lat}$ for the $n=47$ Rydberg state used in this work.

\subsection*{Atom rearrangement for GHZ states}

Performing rearrangement in an optical lattice is crucial for the all-to-all multi-qubit gates presented, enabling small interatomic spacings which allows many atoms to be placed within a single Rydberg blockade. The rearrangement protocol used for this experiment has been described in detail previously~\cite{young2023atomic}. For the data presented in this work, the per-atom rearrangement success rate varies between 85-98\%. Generally, we rearrange the atoms within a GHZ ensemble into a rectangular grid of 2--3 rows and columns with spacings between 1--3 lattice sites in each direction; for details of each $N$, see Extended Data Table~\ref{tabed1}. On each run of the experiment, we prepare $2\times2$, $3 \times 2$ or $3 \times 3$ copies of fixed-size ensembles; for GHZ-state cascades, we prepare the distribution shown in Fig.~\ref{fig4}a. The minimum spacing between ensembles along a single direction is $14a_{\rm lat}$. The maximum GHZ-state size of 9 achieved is limited by the principal Rydberg quantum number of 47 and a few technical limitations on the exact rearrangement patterns we are able to currently prepare. Based on the current trend in measured raw fidelities, resolving these technical challenges might enable preparation of up to 16-atom GHZ states without going to higher-lying Rydberg states.

\subsection*{GHZ-state fidelity measurement correction}
Errors in our state detection scheme can cause the measured GHZ-state fidelity to be different than the true fidelity prepared in the experiment. We stress that the raw parity contrast is generally robust to known effects that could cause an overestimation of the fidelity, and thus the $0.61(1)$ raw parity contrast measured for $N=9$ certifies genuine 9-particle entanglement. Nevertheless, performing measurement correction can help to more accurately assess the preparation fidelity of the GHZ state; we describe the procedure we use here. The measurement-corrected fidelities are shown in Extended Data Table~\ref{tabed2}.

Misidentification of bright sites as dark and vice versa (see State detection) tends to reduce the observed GHZ-state fidelity. To correct these errors, we follow a similar procedure to Ref.~\cite{omran2019generation}. Let $p_{n, \mathrm{raw}}$ denote the measured probability of detecting $n$ atoms in $\ket{1}$, and let $p_{n,\mathrm{true}}$ denote the true probability which we would like to determine. We assume that these probabilities are related by a measurement matrix $M_{mn}$ such that 
\begin{align}
    p_{m, \mathrm{raw}} = \sum_{n=0}^N M_{mn} p_{n,\mathrm{true}}.
\end{align}
$M_{mn}$ describes the probability that a state with $n$ atoms in $\ket{1}$ is detected as having $m$ atoms in $\ket{1}$. When $m \leq n$, we have 
\begin{multline}
    M_{m \leq n,n} = \sum_{k=n-m}^{\mathrm{min} \left(n, N-m \right)} \biggl[ {n \choose k} p_{b \to d}^k (1-p_{b\to d})^{n-k} \times \\ {N-n \choose k-n+m} p_{d \to b}^{k-n+m} (1 - p_{d \to b})^{N-k-m} \biggr].
\end{multline}
 When $m>n$, we instead have
\begin{multline}
    M_{m > n,n} = \sum_{k=m-n}^{\mathrm{min} \left( N-n, m \right)} \biggl[ {N-n \choose k} p_{d \to b}^k (1-p_{d \to b})^{N-n-k} \times \\ {n \choose k-m+n} p_{b \to d}^{k-m+n} (1 - p_{b \to d})^{m-k} \biggr].
\end{multline}
We note that this procedure assumes that the infidelity rates are independent across the atoms in an ensemble. To extract $p_{n, \mathrm{true}}$, we perform numerical minimization of $\sum_{m=0}^N \abs{p_{m, \mathrm{raw}} - \sum_{n=0}^N M_{mn} p_{n,\mathrm{true}}}^2$. This correction is relevant for both the populations and parity oscillation measurements.

An error which can cause the GHZ-state fidelity to be overestimated is leakage out of the computational subspace, which leads to an incorrect association of bright sites with $\ket{1}$ and dark sites with $\ket{0}$. This includes loss from the trap (see Lattice release and recapture) and decay to other states (see Effective state decay). In principle, the inferred GHZ-state populations $p_0 + p_N$ can be increased or decreased due to this; here we are only concerned with correcting for a possible overestimation. To do this, we use the scan of the phase $\alpha_c$ for the $\hat{X}(\pi/2)$ rotation initializing the GHZ state (see GHZ preparation and fidelity). $p_0 + p_N$ oscillates with a period $\pi$ as $\alpha_c$ varies; a discrepancy in this value between the calibrated $\alpha_c$ and $\alpha_c + \pi$ indicates a contribution from states with leakage. We fit the measured populations as a function of $\alpha_c$ to the form 
\begin{align}
    p_0 + p_N = \left[ C - A \sin^2 \left( \frac{\alpha_c - \alpha}{2} \right) \right] f(\alpha_c-\alpha) + y.
\end{align}
Here $C$, $A$, $\alpha$ and $y$ are fit parameters, and $f(\alpha_c)$ is the analytically computed function describing the oscillation in $p_0+p_N$ for a perfect GHZ-state. For $N=6,8,9$, we subtracted off $\abs{A}$ from the GHZ-state populations. For $N=4$, the fit implied that we had measured the populations at the lower value, and thus we did not apply this correction. For $N=2$ where an $\hat{X}(-\pi/4)$ rotation was instead used to initialize the Bell state, we perform an additional $\pi$-pulse to invert the populations to obtain the correction. Since the coherence is inferred from the contrast of the parity oscillation, we expect that it is robust to this error and do not apply a corresponding correction.

\subsection*{Sources of error in GHZ-state preparation}

We perform master equation simulations with stochastic sampling of fluctuating parameters to model the effects of various errors present in the experiment. The result of this model for the Bell state and 4-atom GHZ state protocols are shown in Extended Data Fig.~\ref{fig4ed}b. The simulation includes the ground, clock and Rydberg states, as well as an additional state capturing scattering and decay into and out of \tripptwo. The fidelity is calculated explicitly including the parity rotation in the simulation. We use a $2 a_{\rm lat}$ spacing for the Bell state, and a square arrangement with the same minimum spacing for the 4-atom GHZ state. The release and recapture is not explicitly included in the simulation, though we estimate the recapture loss due to single-photon recoil based on the calculated time spent in the Rydberg state. For both the Bell state and the 4-atom GHZ state, our model is able to account for roughly 1/3 of the observed measurement-corrected infidelity.

There are a number of error sources which are more challenging to accurately characterize, but which we expect might explain a significant fraction of the unaccounted error. Numerically, we find that the multi-qubit gates are significantly more sensitive to variations in the Rydberg Rabi frequency $\Omega_r$ (see Extended Data Fig.~\ref{fig4ed}a); imperfections in our calibration procedure of $\Omega_r$ not only directly cause infidelity, but will also increase the infidelity contribution from shot-to-shot fluctuations or inhomogeneity in $\Omega_r$. In the future, more precise calibration procedures~\cite{evered2023high} as well as robust pulse design~\cite{jandura2023optimizing} could help to mitigate these errors. Transduction of the Rydberg phase modulation to amplitude modulation (see Rydberg excitation section) is another source of uncontrolled error on our gates. This can be straightforwardly mitigated by an additional pass through an AOM to counteract the deflection. Beyond that, a more careful characterization of the laser amplitude and phase profiles for various $\Nmax$ will be necessary to discern potential discrepancies between our model of the Rydberg pulse and the actual experiment, for instance due to sharp jumps in the phase modulation (see $\Nmax=10$ in Fig.~\ref{fig1}b). Finally for the $N \geq 6$ ensembles with certain atoms separated only by a single lattice spacing, it may be important to further understand the degree to which the complicated Rydberg interaction spectrum at small separations affects the dynamics \cite{derevianko2015effects}.

\subsection*{GHZ-state stability in atom-laser comparison}
For the atom-laser comparison, we attempt to prepare $M$-copies of $N$-atom GHZ ensembles on each run of the experiment $q$. Let $j=1,\hdots,M$ index the ensembles on a single shot $q$. Because of imperfect rearrangement, each ensemble will have $N_j^{(q)} \leq N$ atoms; critically, the form of $\UGHZ$ ensures that these partially filled ensembles will still be prepared in a GHZ state. Unfilled ensembles $N_j^{(q)}=0$ are removed, and $M$ is reduced for the shot to only count the number of ensembles with $N_j^{(q)}>0$. During the Ramsey dark time $T$, each GHZ state will accumulate a phase $\theta_j^{(q)} = \int_0^T 2 \pi N_j^{(q)} \delta(t) \dd t$, where $\delta(t)$ is the stochastically varying atom-laser detuning. This phase is converted into a parity measurement by an $\hat{X}(\pi/2)$ rotation, with the phase $\phi_c$ calibrated to be near a zero-crossing of the parity oscillation for all possible ensemble sizes. The measurement yields $M$ binary parity outcomes $\mathcal{P}_{z,j}^{(q)} = \pm 1$ for each ensemble. Taking $\langle \hat{\mathcal{P}}_{z,j}^{(q)} \rangle = C_{N_j^{(q)}} \sin \theta_j^{(q)}$ as the parity expectation model, we use the locally unbiased estimator about $\delta = 0$ to convert the measured $\mathcal{P}_{z,j}^{(q)}$ into a single-shot detuning estimate $\delta_{\rm est}^{(q)} = \frac{1}{M} \sum_{j=1}^M \mathcal{P}_{z,j}^{(q)}/(2 \pi N_j^{(q)} C_{N_j^{(q)}} T)$. Here $C_{N_j^{(q)}}$ is the parity contrast at $t=0$ for an $N_j^{(q)}$-atom GHZ state after application of the $\Nmax = N$ gate. Because we only calibrated the contrast $C_N$ of the maximum GHZ-state size $N$ before these experiments, we used $\abs{C_{N_j^{(q)}}} = \abs{C_N}$ independent of $N_j^{(q)}$; note that this will overestimate the noise and thus provide an upper bound on the reported instability.

A low-bandwidth digital servo converts these detuning estimates into corrections $-\delta_{\rm corr}^{(q)}$, which are used to stabilize the clock laser frequency to the atomic transition. The overlapping Allan deviation is computed for the fractional frequency detuning $y = \delta_{\rm est}/\nu_0$. We use the servo input ($\delta_{\rm est}^{(q)}$) as opposed to the servo output ($-\delta_{\rm corr}^{(q)}$) since the latter is dominated by variations in the magnetic field (see Clock and Rydberg coherence). The exact same procedure and analysis are used for the CSS, where the only change is in the initial assumption where instead $M \times N$-copies of ``1-atom GHZ states" are prepared.

\subsection*{Phase estimator for cascaded GHZ states}

A single measurement of a cascade with $K$ different GHZ-state sizes $N_k$ yields $K$ binomial outcomes $m_k$. $m_k$ describes the number of even parity events (successes) observed out of $M_k$ copies (trials) with probability of success on any single trial $Q_k(\phi) = \left[ 1 + \langle\hat{ \mathcal{P}}_{z,k}(\phi) \rangle \right]/2$. $\langle\hat{ \mathcal{P}}_{z,k}(\phi) \rangle $ is a model of the parity expectation value as a function of $\phi$ which we take to be of the sinusoidal form 
\begin{align}
    \langle\hat{ \mathcal{P}}_{z,k}(\phi) \rangle  = C_k \sin \left[ N_k \left( \phi - \phi_k \right) \right] + y_k.
\end{align}
$C_k$, $\phi_k$ and $y_k$ are model parameters which we fit for in Fig.~\ref{fig4}b. For comparison to an ideal cascade we take $C_k = 1$, $\phi_k = 0$ and $y_k = 0$.

To convert a set $\lbrace m_k \rbrace$ from a single cascade measurement to a phase estimate, we employ the minimum MSE estimator~\cite{demkowicz2015quantum, marciniak2022optimal} defined as follows. The conditional probability of observing $\lbrace m_k \rbrace$ given $\phi$ is 
\begin{align}
    P \left( \lbrace m_k \rbrace | \phi \right) = \prod_{k=1}^K {M_k \choose m_k} \left[ Q_k(\phi) \right]^{m_k} \left[ 1 - Q_k(\phi) \right]^{M_k-m_k}.
    \label{eqpmphi}
\end{align}
The posterior probability can then be computed from Bayes' law using $P\left(\phi | \lbrace m_k \rbrace \right) = P\left( \lbrace m_k \rbrace | \phi \right)\, P(\phi)/P \left(\lbrace m_k \rbrace \right)$. We take the prior knowledge to be a Gaussian of standard deviation $\sigma_{\phi}$
\begin{align}
    P(\phi) = \frac{1}{\sqrt{2\pi}\sigma_{\phi}} \exp\left(- \frac{\phi^2}{2 \sigma_{\phi}^2} \right).
\end{align}
For our proof-of-principle phase estimation experiments performed at $T=0$, we chose $\sigma_{\phi} = \pi/6$ larger than the inversion range of the maximum size $N_{K=4}=8$ GHZ state such that the cascade is required to make unambiguous estimates; $\sigma_{\phi}$ must also not be too large as to possess significant weight beyond the maximum dynamic range $[-\pi,\pi]$ (though additional schemes can be employed to overcome this limitation~\cite{rosenband2013exponential, borregaard2013efficient}). For clock applications, $\sigma_{\phi}$ should be chosen to reflect the spread in integrated atom-laser detuning at the Ramsey dark time being used for interrogation~\cite{kaubruegger2021quantum, marciniak2022optimal}. The marginal distribution $P(\lbrace m_k \rbrace)$ is given by integrating the conditional over the prior $P \left(\lbrace m_k \rbrace \right) = \int_{-\infty}^{\infty} \dd \phi \,  P(\lbrace m_k \rbrace | \phi) \, P(\phi)$. Finally, the minimum MSE estimator is given by 
\begin{align}
    \phi_{\rm est} \left( \lbrace m_k \rbrace \right)  = \int_{-\infty}^{\infty} \dd \phi \, P \left( \phi | \lbrace m_k \rbrace \right) \, \phi.
\end{align}
This estimator then provides a map from any possible outcome set $\lbrace m_k \rbrace$ to real numbers. It can be fully defined once a model $\langle\hat{ \mathcal{P}}_{z,k}(\phi) \rangle $ is given for any $M_k$ and $N_k$.

This performance of this estimator can be evaluated by considering the mean estimate 
\begin{align}
    \bar{\phi}_{\rm est} = \sum_{\lbrace m_k \rbrace} P\left( \lbrace m_k \rbrace | \phi \right) \phi_{\rm est}(\lbrace m_k \rbrace),
\end{align}
and the MSE
\begin{align}
    \Delta \phi_{\rm est}^2 = \sum_{\lbrace m_k \rbrace} P\left( \lbrace m_k \rbrace | \phi \right) \left[ \phi_{\rm est}(\lbrace m_k \rbrace) - \phi \right]^2.
\end{align}
For all theoretical curve in Figs.~\ref{fig4}d-f, $P(\lbrace m_k \rbrace | \phi)$ is obtained using the binomial expression (\ref{eqpmphi}); for the experimental results, it is approximated based on a bootstrap resampled distribution from the data in Fig.~\ref{fig4}b (see Bootstrapping of phase estimation data). Ideally, the MSE $\Delta \phi_{\rm est}^2$ is as small as possible while maintaining unbiased estimates $\bar{\phi}_{\rm est} = \phi$ for as large a range of $\phi$ as possible. Because the cascades considered in this work have $\langle \hat{\mathcal{P}}_{z,k}(\pm \pi) \rangle \approx 0$ for all $k$, large estimation errors are made at the edge of the range $[-\pi,\pi]$. Though these errors do decrease for larger $K$ cascades, it may be possible to more efficiently mitigate this issue by using local clock rotations~\cite{eckner2023realizing,shaw2024multi}.

\subsection*{Effective measurement uncertainty for frequency estimation}

The performance of the cascade for frequency estimation during clock operation, which uses nonzero dark time, can be predicted from the MSE~\cite{kaubruegger2021quantum, marciniak2022optimal}. This is done by associating the distribution of integrated atom-laser detunings, under a noise model at a specific dark time $T$, with a prior knowledge used for Bayesian frequency estimation~\cite{macieszczak2014bayesian,leroux2017line}. For a Gaussian prior of standard deviation $\sigma_{\phi}$, the effective measurement uncertainty on a single cycle of the clock interrogation is
\begin{align}
    \Delta \phi_{\rm eff} = \frac{\Delta \phi_{\rm BMSE}}{\sqrt{1 - \left( \Delta \phi_{\rm BMSE}/\sigma_{\phi} \right)^2}}.
\end{align}
Here, $\Delta \phi_{\rm BMSE}$ is the Bayesian MSE given by
\begin{align}
    \Delta \phi_{\rm BMSE} = \int_{-\infty}^{\infty} \dd \phi P(\phi) \Delta \phi_{\rm est}^2,
\end{align}
which quantifies the performance of the estimator given the prior knowledge. The expected Allan variance reduction relative to SQL is given by $\Delta \phi_{\rm eff}^2 N$, which is shown in Fig.~\ref{fig4}f. By using a noise model to determine a relation between $\sigma_{\phi}$ and $T$, an absolute instability can be computed from $\Delta \phi_{\rm eff}$~\cite{kaubruegger2021quantum, marciniak2022optimal}.

\subsection*{Bootstrapping of phase estimation data}

To explore GHZ-state cascades with a larger number of copies than can be prepared in a single run of the experiment, the distribution $P(\lbrace m_k \rbrace | \phi)$ is obtained by bootstrap resampling of the parity data in Fig.~\ref{fig4}b. The procedure is repeated for each phase $\phi$, so the following protocol applies for a fixed value of $\phi$. On each run of the experiment, ensembles of various sizes are prepared across 8 different locations, and a binary parity outcome is obtained from each; due to imperfect rearrangement, some ensembles will have fewer atoms than intended. To perform the analysis for a cascade with $K$ different sizes $N_k$ (and $M_k$ copies each), we start by collecting the parity outcomes across all experimental repetitions and ensemble locations into $K$ different sets $\lbrace \mathcal{P}_z^{(l)} \rbrace_{k=1}^K$; in the $k$-th set, $l$ indexes each time an ensemble of size $N_k$ was prepared, and the $\mathcal{P}_z^{(l)}$ are the corresponding binary parity outcomes. A single bootstrap outcome $r$ is obtained by drawing $M_k$ random samples from each set $\lbrace \mathcal{P}_z^{(l)} \rbrace_k$; with $m_k^{(r)}$ counting the number of even parity outcomes from the $k$-th sample, the set $\lbrace m_k^{(r)} \rbrace$ is converted using the estimator function $\phi_{\rm est}\left(\lbrace m_k \rbrace \right)$ into a single bootstrap estimate $\phi_{\rm est}^{(r)}$. Repeating this $R=2000$ times, we obtain a distribution of phase estimates from a bootstrapped sampling of $P(\lbrace m_k \rbrace | \phi)$. The mean estimate and MSE are computed as $\bar{\phi}_{\rm est} = \frac{1}{R} \sum_{r=1}^R \phi_{\rm est}^{(r)}$ and $\Delta \phi_{\rm est}^2 = \frac{1}{R} \sum_{r=1}^R \left( \phi_{\rm est}^{(r)}  - \phi \right)^2$.

\subsection*{Scaling of cascade measurement uncertainty}

In Fig.~\ref{fig4}f, a reference line corresponding to $\Delta \phi_{\rm eff} = \pi \sqrt{\ln N_{\rm tot}}/N_{\rm tot}$ is shown. We empirically found that this line captures the scaling of an ideal cascade reasonably well. Here we comment on a couple of theoretical considerations which roughly inform this guide.

This first consideration is that with finite prior information, the standard HL $1/N$ is not saturable asymptotically. Using optimal Bayesian estimation, it has been shown that the asymptotic precision scaling is instead tightly bounded by a $\pi$-corrected HL~\cite{jarzyna2015true,gorecki2020pi}, which is $\pi/N_{\rm tot}$ for the standard spin-1/2 $\hat{\sigma}_z/2$ phase-encoding Hamiltonian. Because this is only asymptotic, the uncertainty of a finite size system can be better than this limit. Nevertheless, comparing the scaling of $\Delta \phi_{\rm eff}$ to a $\pi$-corrected limit is a natural starting point.

The second consideration is that a non-constant correction can arise due to the resource overhead of using smaller GHZ ensembles. It was explicitly shown for a cascaded GHZ clock, using binary estimates up to the largest GHZ size, that the scaling in the optimal number of copies to sufficiently suppress rounding errors leads to a logarithmic correction over the HL~\cite{kessler2014heisenberg,komar2014quantum}. These works considered a restricted distribution of copies, where the number of copies could increase with the number of different GHZ sizes $K$, but was (mostly) fixed across sizes $k$ for a given $K$. In the problem of pure phase estimation, it has been shown that further allowing the number of copies to vary with $k$, specifically such that there are more copies of smaller ensembles, allows the logarithmic overhead to be removed and HL scaling up to a constant overhead to be recovered~\cite{higgins2009demonstrating, berry2009perform}. The theoretical results for an ideal cascade shown in Fig.~\ref{fig4}f suggest that such a linear distribution does not remove the logarithmic correction in the protocol we considered, though there are a number of potentially important differences. One such difference is the application of known phase shifts on each ensemble to perform readout in different measurement bases, which is a technique that has been recently demonstrated in optical clocks~\cite{shaw2024multi,zheng2024reducing}.

\subsection*{Error bars and fitting}
Error bars on populations and parity measurements are 68\% Clopper-Pearson confidence intervals. Error bars on the Allan deviation represent 68\% confidence intervals assuming white phase noise. Fits of the experimental data are done using weighted least squares and error bars on fitted parameters represent one standard deviation fit errors.

\subsection*{Acknowledgements}
\noindent
We acknowledge earlier contributions to the experiment from M.~A. Norcia and N. Schine as well as fruitful discussions with R. Kaubruegger and P. Zoller.
The authors also wish to thank S. Lannig and A.~M. Rey for careful readings of the manuscript and helpful comments.
In addition, we thankfully acknowledge helpful technical discussions and contributions to the clock laser system from A. Aeppli, M. N. Frankel, J. Hur, D. Kedar, S. Lannig, B. Lewis, M. Miklos, W.~R. Milner, Y.~M. Tso, W. Warfield, Z. Hu, Z. Yao. 
This material is based upon work supported by the Army Research Office (W911NF-22-1-0104), the Air Force Office for Scientific Research (FA9550-19-1-0275), the National Science Foundation QLCI (OMA-2016244), the U.S. Department of Energy, Office of Science, the National Quantum Information Science Research Centers, Quantum Systems Accelerator, and the National Institute of Standards and Technology. 
This research also received funding from the European Union’s Horizon 2020 program under the Marie Sklodowska-Curie project 955479 (MOQS), the Horizon Europe program HORIZON-CL4-2021- DIGITALEMERGING-01-30 via the project 101070144 (EuRyQa) and from the French National Research Agency under the Investments of the Future Program project ANR-21-ESRE-0032 (aQCess).
We also acknowledge funding from Lockheed Martin.
A.C. acknowledges support from the NSF Graduate Research Fellowship Program (Grant No. DGE2040434); W.J.E. acknowledges support from the NDSEG Fellowship; N.D.O. acknowledges support from the Alexander von Humboldt Foundation.

\subsection*{Author Contributions}
\noindent
A.C., W.J.E., T.L.Y., A.W.Y., N.D.O. and A.M.K. contributed to the experimental setup, performed the measurements and analyzed the data. S.J. and G.P. conceptualized the multi-qubit gate design. L.Y. and K.K. contributed to the clock laser system under supervision from J.Y. A.M.K. supervised the work. All authors contributed to the manuscript.

\subsection*{Competing Interests}
\noindent
G.P. is co-founder and shareholder of QPerfect.

\subsection*{Materials and Correspondence}
\noindent
Correspondence and requests for materials should be addressed to A.M.K.

\subsection*{Data Availability}
\noindent
The data that support the findings of this study are available from the corresponding
author upon reasonable request. Source data for figures 1--4 are provided with the paper.

\setcounter{figure}{0}
\renewcommand{\figurename}{Extended Data Fig.}

\begin{figure*}
    \centering
    \includegraphics[width = \textwidth]{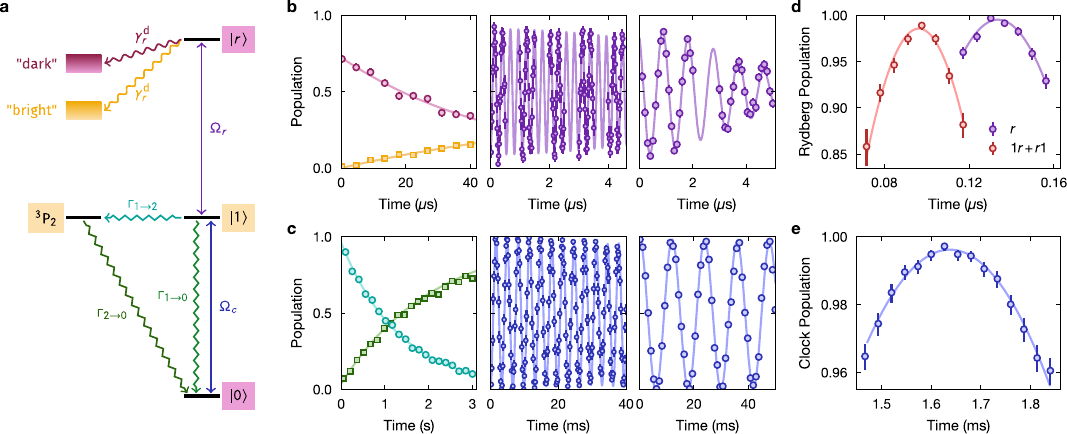}
    \caption{\textbf{Characterizing clock and Rydberg operations.}
    \textbf{a}, Effective level diagram for clock qubits with Rydberg coupling. 
    Wavy lines indicate Rydberg decay. We categorize the many possible Rydberg decay paths by whether the final state is dark (dark pink) or bright (orange) to our standard state detection scheme; note that these final states include the ones that are explicitly shown, with the background color indicating dark or bright. 
    Jagged lines indicate Raman scattering paths (intermediate state not shown). 
    Straight lines indicate coherent drives.
    \textbf{b}, (left) Decay of Rydberg state over time to states dark (dark pink circles) and bright (orange squares) to the detection protocol.
    We fit an exponential $1/e$ decay time to dark (bright) states of $\tau_r^{\rm d} = 1/\gamma_r^{\rm d} = \SI{51(3)}{\micro\second}$ ($\tau_r^{\rm b} = 1/\gamma_r^{\rm b} = \SI{86(3)}{\micro\second}$ ).
    (middle) Single-atom Rydberg Rabi oscillations at $\Omega_r = 2\pi \times \SI{3.7}{\mega\hertz}$ with a fitted $\SI{11(1)}{\micro\second}$ Gaussian $1/e$ decay time. 
    (right) Single-atom Rydberg Ramsey oscillations at a $\SI{1}{\mega\hertz}$ detuning with a fitted $\SI{4.5(2)}{\micro\second}$ Gaussian $1/e$ decay time.
    \textbf{c}, (left) Population of $\ket{1}$ (turquoise circles) and $\ket{0}$ (green squares) over time due to Raman scattering in the lattice. 
    Fitting to a rate model (see Methods) yields scattering rates of $\Gamma_{1 \to 0} = \SI{0.48(1)}{\hertz}$, $\Gamma_{1\to2} = \SI{0.26(2)}{\hertz}$, and $\Gamma_{2\to0} = \SI{0.47(3)}{\hertz}$ in an $\approx \SI{920}{E_r}$ deep lattice 2D lattice.
    (middle) Clock Rabi oscillations at $\Omega_c = 2 \pi \times \SI{0.31}{kHz}$ yielding a fitted ground state fraction of $0.96(1)$.
    (right) Clock Ramsey oscillations at an $\SI{84}{\hertz}$ detuning with a fitted $\SI{217(17)}{\milli\second}$ Gaussian $1/e$ decay time.
    We note that the longer coherence time reported in Fig.~\ref{fig2}d is obtained by a different method in which the Ramsey fringe contrast is carefully measured at each dark time and out to significantly longer times.
    \textbf{d}, Rydberg $\pi$-pulse fidelity for single atoms (purple) and two-atom blockade (red). These data are SPAM-corrected (see Methods and Ref.~\cite{madjarov2020high}). Parabolic fits yield SPAM-corrected Rydberg $\pi$-pulse fidelities of $\Frydspam$ for single atoms and $\Frydbspam$ for two-atom blockade.
    \textbf{e}, Clock $\pi$-pulse fidelity. A parabolic fit yields a raw clock $\pi$-pulse fidelity of $\Fclock$.
    } 
    \label{fig2ed}
\end{figure*}

\begin{figure*}
    \centering
    \includegraphics[width = \columnwidth]{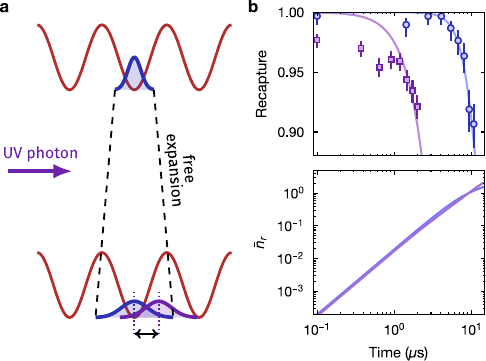}
    \caption{\textbf{Release and recapture in optical lattices.}
    \textbf{a}, Schematic of the release and recapture process.  
    The atoms expand from an initially well-localized state while the lattice is off, and when excited to the Rydberg state, the atoms will additionally undergo a center of mass displacement due to the momentum recoil of the UV photon.
    When the lattice is turned back on, the wavefunction will be projected both into higher band Wannier orbitals as well as nearby sites, causing both loss and heating.
    \textbf{b}, (top) Measured survival as a function of time that the lattice is turned off for the ground state (blue circles) and Rydberg state (purple). 
    The solid lines are theoretical predictions for the recapture probability from an approximately $50 E_r$ lattice (see Methods).
    At short times, the Rydberg-state survival decreases quadratically, and we fit a Gaussian $1/e$ decay time of $\SI{8.7(1)}{\micro\second}$.
    (bottom) Theoretically predicted increase in mean phonon number (see Methods) for recaptured atoms over the same duration. 
    The heating is quadratic at short times, but begins to taper off as the highest energy atoms are lost. 
    The lattice turn-off duration is $<\SI{2}{\micro\second}$ for all data shown in the main text.} 
    \label{fig3ed}
\end{figure*}

\begin{figure*}
    \centering
    \includegraphics{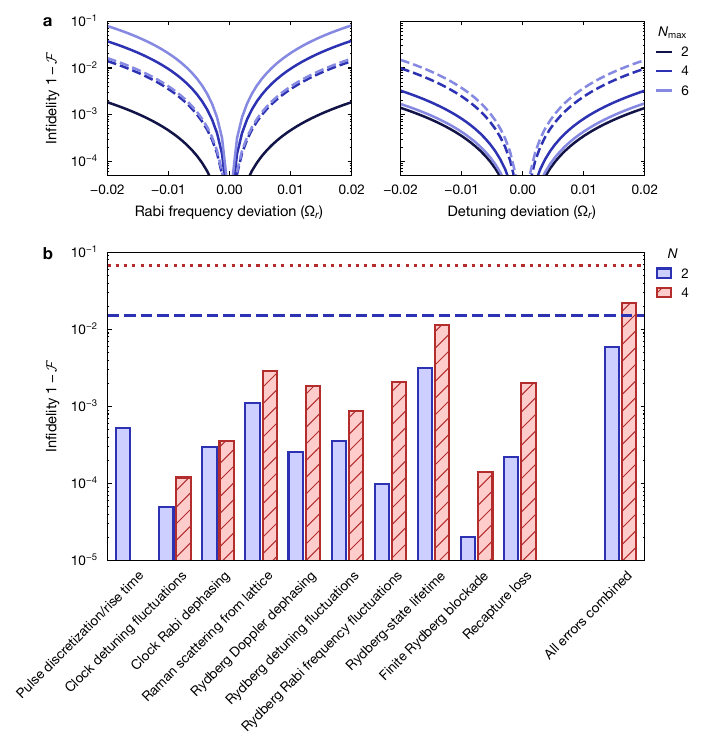}
    \caption{\textbf{Error modeling for GHZ-state fidelities.}
    \textbf{a}, Sensitivity of multi-qubit gate $\UGHZ$ to Rydberg Rabi frequency and detuning deviations for various $\Nmax$. Solid lines indicate infidelity for $N=\Nmax$ GHZ state; dashed lines indicate infidelity for $N=2$ Bell state.  
    \textbf{b}, Modeling of various error sources for $N=2$ Bell state (blue) and $N=4$ GHZ state (red, hatched). For the Bell state, we consider the CZ gate protocol shown in Fig.~\ref{fig1}c; for the 4-atom GHZ state, we consider the general $\Nmax = N$ scheme shown in Fig.~\ref{fig2}a. The measurement corrected Bell state [4-atom GHZ state] infidelity (see Extended Data Table~\ref{tabed2}) is shown as the blue, dashed [red, dotted] line. In both cases, our error model accounts for roughy 1/3 of the observed infidelity. The presence of pulse discretization/rise time error for only the Bell state is because we use the exact time-optimal CZ gate implementation described in Ref.~\cite{evered2023high} as opposed to a modulation optimized for our pulse model.
    } 
    \label{fig4ed}
\end{figure*}

\clearpage

\setcounter{table}{0}
\renewcommand{\tablename}{Extended Data Table}
\begin{table*}
    \setlength{\tabcolsep}{10pt}
    \centering
    \begin{tabular}{*6c}
        \toprule
        N & $\Delta x$ & $\Delta y$ & $l_y$ & $l_x$ & $U_{\rm min}$ \\
         & $(a_{\rm lat})$ & $(a_{\rm lat})$ & (rows) & (columns) & ($\hbar \Omega_r$) \\
        \midrule
        2 & 2 or 3 & N/A & 1 & 2 & 99(2) \\
        4 & 2 or 3 & 2 & 2 & 2 & 32.7(6) \\
        6 & 3 & 1 & 3 & 2 & 32.7(6) \\
        8 or 9 & 2 & 1 & 3 & 3 & 9.0(2) \\
        \bottomrule
    \end{tabular}
    \caption{
    \textbf{Atomic arrangement parameters for different GHZ ensemble sizes.}
    All patterns are oriented in a rectangular pattern of $l_y$ rows by $l_x$ columns on the square lattice, with spacings $\Delta x$ and $\Delta y$ along each direction. The minimum blockade is computed as $U_{\rm min} = C_6/r_{\rm max}^6$, where $r_{\rm max}^6 = \sqrt{(l_x \Delta x)^2 + (l_y \Delta y)^2}$ and $C_6 = 2 \pi \times \SI{10.4(2)}{\giga \hertz \cdot \micro \meter^6}$ is obtained from measurements of the transition frequency for two-photon excitation of $\ket{11} \to \ket{rr}$; we note that this $C_6$ value is roughly 15\% larger than we reported previously in Ref.~\cite{eckner2023realizing}. $U_{\rm min}$ is given in units of $\Omega_r = 2\pi \times \SI{4}{\mega \hertz}$, even though the actually Rabi frequency used in various experiments varies between 3--4\,MHz. The $N=8$ data used the same pattern as $N=9$, but with a single corner atom removed.}
    \label{tabed1}
\end{table*}

\begin{table*}
    \setlength{\tabcolsep}{10pt}
    \centering
    \begin{tabular}{*7c}
        \toprule
          & \multicolumn{3}{c}{Raw} & \multicolumn{3}{c}{Measurement-corrected} \\ \cmidrule(r){2-4} \cmidrule(l){5-7}
         $N$ & $p_0 + p_N$ & $C$ & $\mathcal{F}$ & $p_0 + p_N$ & $C$ & $\mathcal{F}$ \\  
        \midrule
        2 & 0.990(2) & 0.975(3) & 0.983(2) & 0.988(4) & 0.983(3) & 0.985(2) \\
        4 & 0.940(7) & 0.88(2) & 0.912(8) & 0.955(7) & 0.91(2) & 0.933(8) \\
        6 & 0.908(6) & 0.77(1) & 0.837(6) & 0.90(5) & 0.82(1) & 0.86(2) \\
        8 & 0.822(8) & 0.68(1) & 0.750(7) & 0.77(6) & 0.75(1) & 0.76(3) \\
        9 & 0.80(1) & 0.61(1) & 0.707(9) & 0.8(1) & 0.68(1) & 0.75(5) \\
        \bottomrule
    \end{tabular}
    \caption{
    \textbf{Summary of raw and measurement-corrected GHZ-state fidelities.}
    The measured values of the GHZ-state populations $p_0 + p_N$, parity oscillation contrast $C$ and GHZ-state fidelity $\mathcal{F}$ are shown for both the raw data and after applying measurement correction (see Methods) for varying size $N$.}
    \label{tabed2}
\end{table*}

\end{document}